\documentclass[numberedappendix,apj,twocolumn]{emulateapj}

\newcommand{\vi}{$V_{606} - i_{775}$}
\newcommand{\iz}{$i_{775} - z_{850}$}
\newcommand{\lya}{Ly$\alpha$}
\newcommand{\vdrop}{$V_{606}$-dropout }
\newcommand{\vdrops}{$V_{606}$-dropouts }

\newcommand{\bFilter}{$B_{435}$ }
\newcommand{\vFilter}{$V_{606}$ }
\newcommand{\iFilter}{$i_{775}$ }
\newcommand{\zFilter}{$z_{850}$ }

\newcommand{\bF}{$B_{435}$}
\newcommand{\iF}{$i_{775}$}
\newcommand{\zF}{$z_{850}$}
\newcommand{\vF}{$V_{606}$}

\shorttitle{UDF05: Faint End of the LBG LF at $z \sim 5$}
\shortauthors{Oesch et al.}

\begin{document}

\title{The UDF05 Follow-up of the HUDF: I. The Faint-End Slope of the Lyman-Break Galaxy Population at $z \sim 5$ 
\altaffilmark{1}}

\altaffiltext{1}{Based on data obtained with the Hubble Space Telescope operated by AURA, Inc. for NASA under contract NAS5-26555}

\author{P. A. Oesch\altaffilmark{2}, 
M. Stiavelli\altaffilmark{3,4}, 
C. M. Carollo\altaffilmark{2}, 
L. E. Bergeron\altaffilmark{3}, 
A. M. Koekemoer\altaffilmark{3},
R. A. Lucas\altaffilmark{3}, 
C. M. Pavlovsky\altaffilmark{3}, 
M. Trenti\altaffilmark{3}, 
S. J. Lilly\altaffilmark{2}, 
S. V. W. Beckwith\altaffilmark{3},
T. Dahlen\altaffilmark{3},
H. C. Ferguson\altaffilmark{3},
Jonathan P. Gardner\altaffilmark{5},
C. Lacey\altaffilmark{6},
B. Mobasher\altaffilmark{3},
N. Panagia\altaffilmark{3,7,8},
H.-W. Rix\altaffilmark{9}
}

\altaffiltext{2}{Institute of Astrophysics, ETH Zurich, CH - 8093 Zurich, Switzerland; poesch@phys.ethz.ch}
\altaffiltext{3}{Space Telescope Science Institute, Baltimore, MD 21218, United States}
\altaffiltext{4}{Department of Physics and Astronomy, Johns Hopkins University, Baltimore, MD 21218, United States}
\altaffiltext{5}{Laboratory for Observational Cosmology, Code 665, NASA's Goddard Space
Flight Center, Greenbelt MD 20771}
\altaffiltext{6}{Institute for Computational Cosmology, Department of Physics, University of Durham, South Road, Durham, DH1 3LE, UK}
\altaffiltext{7}{INAF- Osservatorio Astrofisico di Catania, Via S. Sofia 78, I-95123 Catania, Italy}
\altaffiltext{8}{Supernova Ltd., OYV 131, Northsound Road, Virgin Gorda, British Virgin Islands}
\altaffiltext{9}{Max-Planck-Institute for Astronomy, D - 69117 Heidelberg, Germany}

\begin{abstract}
We present the UDF05 project, a HST Large Program of deep ACS (F606W, F775W, F850LP) and NICMOS (F110W, F160W) imaging of three fields, two of which coincide with the NICP1-4 NICMOS parallel observations of the Hubble Ultra Deep Field (HUDF). In this first paper we use the ACS data for the NICP12 field, as well as the original HUDF ACS data, to measure the UV Luminosity Function (LF) of $z \sim 5$ Lyman Break Galaxies (LBGs) down to very faint levels. Specifically, based on a $V-i$, $i-z$ selection criterion, we identify a sample of 101 and 133 candidate $z\sim 5$ galaxies down to $z_{850}=28.5$ and 29.25 magnitudes in the NICP12 and in the HUDF fields, respectively. Using an extensive set of Monte Carlo simulations we derive corrections for observational biases and selection effects, and construct the rest-frame 1400 \AA\ LBG LF over the range $M_{1400}=[-21.4, -17.1]$, i.e. down to $\sim 0.04 \ L_{*}$ at $z\sim5$, and complement it with data from the Subaru Deep Field (SDF) from \citet{yosh06} to extend it to the brighter end ($M_{1400}\ge-22.2$). We show that: (i) Different assumptions regarding the SED distribution of the LBG population, dust properties and intergalactic absorption result in a 25\% variation in the number density of LBGs at $z\sim5$; (ii) Under consistent assumptions for dust properties and intergalactic absorption, the HUDF is about 30\% under-dense in $z\sim5$ LBGs relative to the NICP12 field, a variation which is well explained by cosmic variance; (iii) The faint-end slope of the LF is independent of the specific assumptions for the input physical parameters, and has a value of $\alpha \sim -1.6$, similar to the faint-end slope of the LF that has been measured for LBGs at $z\sim3$ and $z\sim6$. Our study therefore supports no variation in the faint-end of the LBG LF over the whole redshift range $z\sim3$ to $z\sim6$. 
Based on a comparison with semi-analytical models, we speculate that the $z\sim5$ LBGs might have a top-heavy IMF.

\end{abstract}


\section{Introduction}

Deep imaging with the Advanced Camera for Surveys (ACS) and the NICMOS camera onboard the Hubble Space Telescope (HST) has been instrumental in pushing the study of galaxy populations out to the reionization frontier of $z\sim6$ and beyond. The development of the Lyman Break technique has been a milestone for the study of high redshift galaxies, as it allows to identify, from broad-band photometry alone, large numbers of star-forming, but otherwise normal, galaxies at early epochs: star forming galaxies at a given high redshift are identified from the drop in flux by, typically, 1-2 magnitudes, blueward of the detection passband, caused by the Lyman continuum break and the Lyman series blanketing induced by intergalactic neutral hydrogen clouds. While the conceptual idea was worked out already 30 years ago \citep{meie76}, it was only in the 1990's that progress in instrumentation made it possible to identify high-$z$ galaxies using this technique \citep[e.g.][]{guha90,stei92}. Several spectroscopic follow-on studies of LBG candidates have proven the high efficiency and reliability of the Lyman Break technique in identifying galaxies at early epochs \citep[e.g.][]{stei99,malh05,vanz06}. 

Despite the availability of large samples of high redshift galaxies with low contamination from interlopers, the derivation of a Luminosity Function (LF), and particularly of its faint-end slope, remains difficult on account of the small volume probed by currently available ultra deep surveys such as the Hubble Ultra Deep Field \citep[HUDF;][]{ beck06}, and because of the many corrections which are often not well constrained. This is illustrated by the large scatter in the published LF parameters that have been derived from the same data sets \citep[e.g.][]{Bunker04, Windhorst04, bouw05,beck06}. Mapping the evolution of the faint-end slope of the LF at high redshifts remains a major goal of observational cosmology, as the faint-end slope is expected to be dramatically affected by reionization \citep[e.g.][]{WitheLoeb06}, and a solid detection of a change in slope would be strong evidence for the transition between a partially and a fully reionized universe. As also highlighted by e.g., \citet{beck06}, one important step to improve the sensitivity to changes in the faint end slope is to derive LFs at different redshifts in the most homogeneous way, applying similar techniques to similar data sets.

\begin{figure}
\plotone{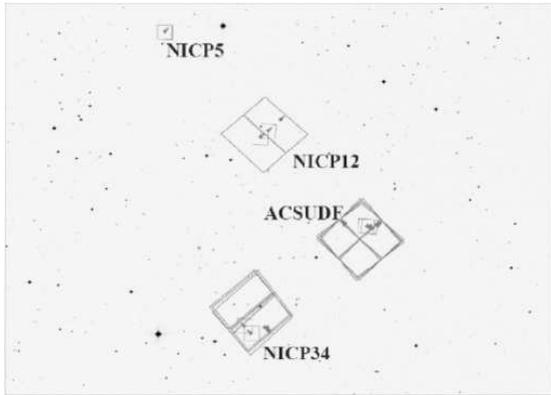}
\caption{The planned UDF05 fields relative to the original HUDF ACS and NICMOS-parallel pointings. The larger and smaller squares correspond to the ACS and NICMOS pointings, respectively. The \vdrop study that we present in this paper is based on the UDF05 ACS observations covering the NICP12 field, and the HUDF. Due to the two-gyro operation of the HST, the UDF05 adds a new deep-imaging NICMOS field to the original HUDF pointings (NICP5) . 
\label{fig:pointing}}
\end{figure}

This paper presents the UDF05 project, a 204-orbit HST Large Program of ultradeep ACS and NICMOS observations of multiple fields to study the evolution of the faint-end of the LF throughout the $z\sim4-8$ redshift regime. The UDF05 was originally constructed to observe with ACS/WFC the two NICMOS parallel fields that were acquired while the HUDF was imaged with the ACS. These HUDF-NICMOS parallel fields (hereafter NICP12 and NICP34, see Fig. \ref{fig:pointing}), have a total exposure of $\sim 2 \times 10^5$ s both in the F110W (broad J) and F160W (H) filters, providing the deepest near-IR data available. Only relatively shallow ACS data were available at their locations, and our additional UDF05 data for these fields was aimed at reaching a depth in the visible, comparable to that of the NICMOS observations (see Table 1). Furthermore, the planned UDF05 pointings should have had an orientation such that, while acquiring the ACS data for the NICP12 and NICP34 fields, parallel NICMOS observations should have been taken for the original HUDF, so as to obtain a total of three fields with ultradeep imaging both in the visible and in the near-IR. Unfortunately, the transition of HST to the two-gyro mode and the first failure of ACS in June 2006 severely limited the number of available orbits at the required orientation, so that the NICMOS observations of the original ACS HUDF field ended up with only one third of the exposure time originally planned.

In our first utilization of the UDF05 dataset we study the faint-end of the LF of $z\sim5$ $V_{606}$-dropout galaxies; specifically, we compare the $z\sim5$ LF obtained from the ACS NICP12 field with that derived from the original ACS HUDF, in order to assess the effects of cosmic variance on the $z\sim5$ LF parameters. In Oesch et al.\ (2007, in preparation) and Stiavelli et al.\ (2007, in preparation) we will extend our 
exploration of the realistic uncertainties in the faint-end slope of the LF of Lyman Break Galaxies to the $z\sim6$ and $z\geq7$ regimes, respectively, probing well into the expected reionization epoch and thus constraining its impact on galaxy formation.

This paper is structured as follows. After the description of the data, corrections for noise correlation and source detection (\S \ref{sec:data}), we describe our selection criteria (\S \ref{sec:colselection}), and the simulations that were performed to estimate the completeness of our source catalogs (\S \ref{sec:simulations}). The observed surface densities are derived by including the effects of photometric errors (\S \ref{sec:vdropsources}) and the LF is computed under different assumptions about the underlying SED distribution of the LBG population as well as different prescriptions for intergalactic hydrogen absorption (\S \ref{sec:LF}). In section \S \ref{sec:cosmicscatter} we present a theoretical estimate for the effects of cosmic variance on our results. In section \S \ref{sec:discussion} we discuss the evolution of the faint-end of the LF of star forming galaxies in the $z\sim3-6$ redshift regime, and compare our observed $z\sim5$ LF with theoretical predictions based on a standard\footnote{Throughout this paper we adopt the concordance cosmology: $\Omega_M=0.3, \Omega_\Lambda=0.7, H_0=70$ kms$^{-1}$Mpc$^{-1}$, i.e. $h=0.7$. Magnitudes are given in the AB system \citep{okeg83}.} $\Lambda$ Cold Dark Matter ($\Lambda$-CDM) universe. Section \S \ref{sec:conclusion} summarizes our main results and conclusions. A comparison of our LF measurements with published observational estimates is shown in Appendix \ref{sec:comparison}.

\section{Data}
\label{sec:data}

The HUDF ACS WFC data are the same as described by \citet{beck06} and include exposures in the F435W (hereafter \bFilter), F606W (hereafter \vF), F775W (hereafter \iF), and F850LP (hereafter \zF) filters for a total of 400 orbits. We used the publicly released HUDF images without further processing. 

As a part of the UDF05 project, ACS WFC data were obtained for the NICP12 field in the F606W, F775W, and F850LP filters amounted to a total of 101 HST orbits; Table \ref{tab:obslog} lists the specifics of these observations. The center of the NICP12 field is located at RA(2000)=03:33:03.60, Dec(2000)=$-$27:41:01.80; Figure \ref{fig:pointing} shows all UDF05 pointings relative to those of HUDF ACS and NICMOS-parallels. The UDF05 ACS observations of the NICP12 field employ a larger dithering step than those adopted for the original HUDF; this choice was made in order to obtain an improved sub-pixel sampling for the NICMOS parallel data that were acquired simultaneously, relative to the original HUDF NICMOS-parallel observations.

\begin{deluxetable*}{ccccc}
\tablecaption{Properties of the UDF05 and HUDF Observations \label{tab:obslog}}
\tablewidth{0 pt}
\tablecolumns{5}
\tablehead{\colhead{Field} & \colhead{Filter} & \colhead{HST Orbits} & \colhead{Exposure Time [s]} & \colhead{$10 \sigma$-magnitude\tablenotemark{a} [AB] }}

\startdata
NICP12 & $V_{606}$ & 	9  & 21'600	&	28.49 \\
NICP12 & $i_{775}$ &	23 & 54'000	&	28.44 \\
NICP12& $z_{850}$ &	69\tablenotemark{b} & 168'000 	&	28.47 \\
&           &      &         &         \\
HUDF & $B_{435}$ & 56 & 134'900 & 29.38 \\
HUDF & $V_{606}$ & 56 & 135'300 & 29.81 \\
HUDF & $i_{775}$ & 144 & 347'100 & 29.43 \\
HUDF & $z_{850}$ & 144 & 346'600 & 28.73 
\enddata

\tablenotetext{a}{within an aperture of 0\farcs15 radius}
\tablenotetext{b}{70 planned; one lost due to loss of lock}
\end{deluxetable*}

The exposure times for the UDF05 observations were designed so as to achieve a constant AB-magnitude depth in all filters for all fields. The individual $\sim$1200s exposures (two in each orbit) were combined and re-pixeled to a scale of 30 mas using the task "multidrizzle"; the task "tweakshifts" was used to optimize the shifts between different images. Superdarks and superbiases were obtained combining several hundred single frames to increase the S/N. Finally, a herringbone artifact was removed from the data by processing the images with a Fourier filter, and the electronic ghost present in the ACS images was eliminated by means of an empirical model. The model is built upon the fact that the electronic ghost of a source is located in quadrants different from the one containing the source and on the fact that ACS/WFC is read out from each corner so that if one dithers a source closer to its readout corner the ghost will move in the opposite direction closer to its own corner. Thus, when we drizzle together a dithered pair of images, we align the sources, and smooth out the ghosts. Our algorithm (conceived by L.E. Bergeron) is based on the idea of flipping each quadrant before drizzling. This simple step coadds the ghosts and smooths out the sources and, when applied iteratively with proper masking, can be used to obtain clean images of the ghosts which are then subtracted from the images. These last two steps were developed after the first public release of the UDF05 data (version 1, "v1"). The newly processed images of the UDF05 that we use in our analysis represent an improved "version 2" (v2) of the final products. 

The PSF FWHM of the UDF05 final ACS images are about 0\farcs1 in all passbands.
In the UDF05 NICP12 images, the 10 $\sigma$ limiting magnitudes in apertures of 0\farcs15 radius are 28.49, 28.44, and 28.47 respectively in \vF, \iF, and \zF\  (see Table \ref{tab:obslog}), i.e. the planned uniform sensitivity in all passbands is achieved. For the HUDF the \bF, \vF, \iF, and \zF\ 10 $\sigma$ limiting magnitudes in apertures of 0\farcs15 radius are 29.38, 29.81, 29.43, 28.73 respectively.

\subsection{Noise Correlation}

A point kernel was used for all passbands when drizzling the data, with the exception of the NICP12 \vFilter image, for which the pixels were not shrunken to a point before drizzling to the output frame (pixfrac $\neq0$), as the dithering pattern was inadequate to properly sample the pixels. As a result, this image is expected to show significant noise correlation. 
However, the noise level measured by computing the rms of cleaned, block averaged areas of the \vFilter images, or by computing the total flux in the auto-correlation peak shows a higher degree of correlation than the one due to dithering which is well characterized by the model published by \citet{case00}. Moreover, some noise correlation is also measured in the \iFilter and \zFilter images which were not expected to have any. The unexpected correlated noise seen in these images is of the same amplitude as the additional component seen in the \vFilter image and we believe that this residual noise correlation is introduced by the reference files (flat fields, biases, darks) and, possibly, by unresolved background sources. Any feature present in a given pixel in the reference files is propagated to a larger area on the final science images by the adopted dithering pattern. Neglecting this noise component would lead to overestimating the depth achieved by the observations; for example, in the NICP12 field, the noise would be underestimated by 30\% for galaxies with an area of 0.3 $\times$ 0.3 arcsec$^2$. 

Relative to our final ACS UDF05 images, the released ACS HUDF images, used in this paper, are more severely affected by correlated noise, due to presence of the uncorrected herringbone noise and electronic ghost in ACS (as these effects had not been characterized at the time). This explains why, despite the factor of about two in exposure time, the HUDF \zF\ image is only 0.26 mag deeper than our NICP12 \zF\  image. 

The rms maps produced by multidrizzle were rescaled in order to match the expected flux errors of galaxies with a measured area of 80 pixels, which corresponds to the median size of our dropout galaxies. No simple renormalization is optimal for all source sizes. The adopted choice underestimates the S/N for point sources and overestimates it for sources that are more extended than the area considered; however, it provides on average, an optimal renormalization for the typical galaxies we study in this paper. Simulations are used to assess proper errors as a function of the sources' physical parameters.

\subsection{Source Detection}

Source detection and photometry measurements were done with the software tool SExtractor \citep{bert96}, which was run in double image mode with \zFilter as the "detection band". Despite the fact that the \iFilter images for both the HUDF and the NICP12 data reached a similar, if not deeper, magnitude limit than the \zFilter band, the latter was preferred for the source detection because of the larger uncertainties in an \iF-selected LF, which are driven by redshifting of the \lya\ forest into the \iFilter passband at $z \ga 4.8$, significantly decreasing the observed fluxes. 
 
The SExtractor parameters were optimized so as to maximize the number of detected galaxies while minimizing the number of spurious sources. For NICP12, the detection threshold was set to be a minimum of 10 connected pixels 0.55 $\sigma$ above the background. This led to 5,734 objects within the central, highest S/N 11.9 arcmin$^2$ of the \zFilter NICP12 image. Similar SExtractor configurations returned 6,336 objects in the central 11.2 arcmin$^2$ of the HUDF \zFilter image.

In order to estimate the contamination of spurious detections in the source catalogs, we cleaned the images from the detected sources, and reran SExtractor using the same configuration parameters on the "inverted" (i.e., multiplied by -1) images. This contamination was found to be $\la 0.5\%$. 
A final visual inspection was performed to clean the catalogs of residual spurious sources and stellar diffraction spikes. The source magnitudes were corrected for the small amount of Galactic extinction expected in the field, $E(B-V) = 0.007$ and 0.008 in NICP12 and HUDF, respectively \citep{schl98}.

Colour measurements were derived from the galaxy ISO magnitudes (apertures matched to the detection isophotes), since these are closely matched to the higher S/N parts of the objects and, as shown by our simulations (Section \ref{sec:simulations}), include much less noise than the SExtractor AUTO magnitudes which we use for the total magnitude of a source in the detection band. This is especially important for the very compact and faint sources that we study in this paper.

\section{The Selection of $z \sim 5$ Lyman-Break Galaxies}
\label{sec:colselection}

\subsection{The Color-Color Selection Criterion}

The redshift evolution of synthetic galaxy models, properly attenuated by intergalactic hydrogen absorption, can be used to identify regions in a color-color diagram which efficiently disentangle galaxies in a specific high redshift window from lower-redshift interlopers.

\begin{figure}
\plotone{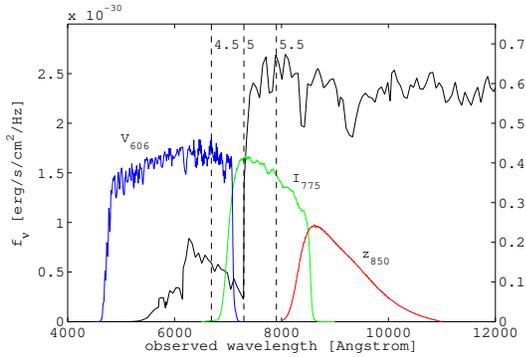}
\caption{Synthetic SED of a starbursting galaxy at $z=5$ with $z_{850}=25.5$ mag. The optical passbands used for the UDF05 program are overplotted. The vertical dashed lines indicate the \lya\ edge at redshifts 4.5, 5 and 5.5, respectively. \label{fig:LBG_z5}}
\end{figure}

For the identification of a \vdrop sample (see Figure \ref{fig:LBG_z5}), this selection window has been optimized to be \citep[see][]{giav04b,beck06}:

\begin{eqnarray}
& V_{606} - i_{775} > \min \left[2, 1.5 + 0.9 \cdot (i_{775} - z_{850}) \right] & \label{eq:selection1}\\ 
& V_{606} - i_{775} > 1.2 & \\
& i_{775} - z_{850} < 1.3. & 
\label{eq:selection3}
\end{eqnarray}
Galaxies in the range $ z \sim 4.5 - 5.7$ are well identified by this color-color criterion; the reliability of the method has been shown to be very efficient ($\sim 90\%$) by spectroscopic follow-up surveys \citep{vanz06}.

The elimination of low redshift interlopers is facilitated by the availability of the \bFilter passband for the HUDF sample.  This is however not available for the NICP12 data. We therefore used the HUDF data to check the impact of including or excluding the additional S/N(\bF)$<3$ criterion that was adopted by \citet{giav04b} and \citet{beck06} in order to minimize the contamination from low-$z$ interlopers. Only one object was rejected from our HUDF dropouts catalog on the basis of this \bF-band constraint. Therefore, for uniformity with the analysis of the NICP12 field, for which the \bFilter is not available, we did not include any \bF-band constraint in our selection of the HUDF \vFilter-dropouts sample.

\begin{figure}
\plotone{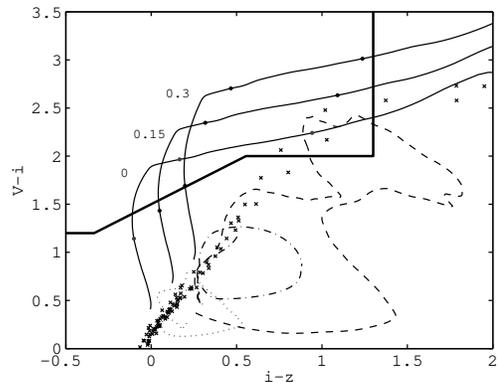}
\caption{Evolutionary tracks of different types of galaxies, and location of the stellar sequence, in the  $V_{606}-i_{750}$ versus $i_{775}-z_{850}$ color-color diagram. The tracks are constructed using the Bruzual \& Charlot (2003) population synthesis models. The thick black line corresponds to the adopted selection (Eq. \ref{eq:selection1} - \ref{eq:selection3}). The thin solid lines show the tracks of 200 Myr old, continuously star-forming galaxies at $z\geq4$ with $E(B-V)=0, 0.15, 0.3$ (indicated in the figure) for a \citet{calz00} dust extinction relation. The redshifts steps 4.5, 5, and 5.5 are indicated with small dots.
Lower redshift galaxy types are plotted with dotted, dash-dotted and dashed lines ($z=0-4$), corresponding to local irregulars, Sbc spirals, and a single stellar population with an age of 3.5 Gyr, respectively. The latter was chosen because it corresponds to a maximally old stellar population at $z\sim1.5$, the redshift at which single-burst early-type galaxies contaminate the selection. The small crosses correspond to the 131 Galactic stars from the \citet{pick98} library.
 \label{fig:tracks}}
\end{figure}

The stellar library of \citet{pick98} shows that stars are expected to lie within a very well defined region on the \vi, \iz\ diagram. Indeed, the few obvious stars in our field lie within this area. To further optimize the removal of stars from our catalogs, the SExtractor stellarity parameter was also considered. This parameter is determined by application of a neural network approach to estimate the likelihood that an object is a point source. {However, this method} is not very reliable at faint magnitudes, and therefore cannot be used to identify stars close to the magnitude limits. 
Inspection of the \vi\ vs. \iz\ diagram and of the half-light radius ($r_{1/2}$) vs. \zFilter diagram showed that sources with stellarity $ > 0.85$, \zFilter $ < 27$ and $r_{1/2} < 0\farcs1$ are most likely stars in the NICP12;  hence all such objects were excluded from our analysis. A similar criterion was applied to the HUDF for \zFilter $ < 28$. Only two and one of such sources were within our \vdrop selection window for the NICP12 and HUDF samples\footnote{All other sources in the HUDF stellar catalog of \citet{pirzkal05} that we do not identify as stars lie outside and very far from the border of our color-color selection window.}, respectively; these objects however had already been visually identified as stars. Although unlikely, stars fainter than \zFilter $ \sim 27$ and 28 may remain a (small) source of contamination in our NICP12 and HUDF catalogs, respectively. 

In Fig. \ref{fig:tracks} the evolutionary tracks of different galaxy types are shown in the $V_{606}-i_{750}$ versus $i_{775}-z_{850}$  color-color diagram. The tracks are built using the population synthesis models of Bruzual \& Charlot (2003). The attenuation by intergalactic hydrogen causes essentially all the galaxy types at redshifts above $\sim$4.5 to enter the selection window. Elliptical galaxies at $z \sim 1.5$ are a source of contamination in the selection criterion due to their red colors; this contamination is however expected to be small in our \vdrop samples, given the expected low surface densities of these galaxies. In fact, none of the seven spectroscopically identified passively evolving galaxies at $z>1.4$ in the HUDF from \citet{dadd06} were selected as \vF-dropouts.

\subsection{Contamination from Low-Redshift Interlopers}

We have estimated the possible fraction of interlopers by applying our selection criterion of Eqs. (1)-(3) to a library of $\sim3000$ synthetic SEDs built upon Bruzual-Charlot (2003) models, adopting the LF derived by \citet{stei99} at $z\sim3$ and no evolution. The models include the effects of intergalactic absorption \citep{mada95}, and span a wide range of metallicities ($0.04-2.5 Z_\odot$), dust reddening, emission lines, and different star formation histories. These included simple stellar populations, models with continuous star formation at constant metallicity, models with self-consistent enrichment during the continuous star formation (estimated with a closed box model and with a model with infall of zero-metallicity gas), and two-burst models combining with five different mass ratios an old ($> 50$ Myr) stellar population with a young ($1-10$ Myr) star forming population that includes self-consistent nebular line and continuum emission.  For each model, versions with and without dust reddening were computed; eight values of reddening were implemented assuming a galactic dust screen with $A_V$ logarithmically spaced between 0.05 and 6.4. Emission lines and nebular continuum were computed in the approximation of case-B recombination for Hydrogen and Helium, based on the ionizing flux from the stellar SED; Oxygen and Carbon line intensities were derived from an analytical interpolation of a grid of calculations performed with the code {\it CLOUDY} \citep{ferla98}. Gas temperatures were derived as a function of metallicity.

\begin{figure}
\plotone{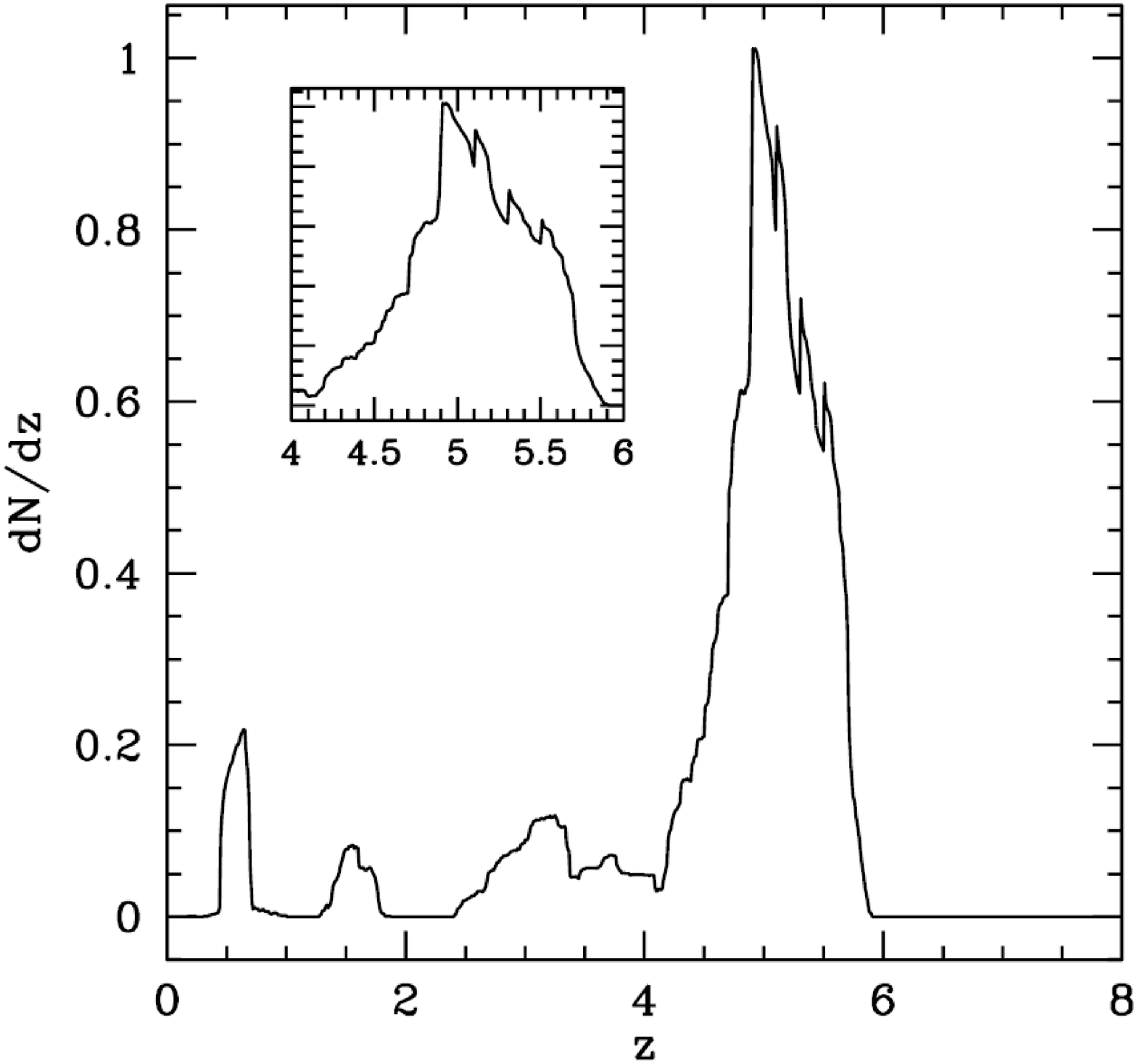}
\caption{The predicted redshift distribution for \vdrop-dropouts as derived assuming a non-evolving LF and synthetic SEDs (see text). The total interloper fraction is estimated to be 23\% and is primarily contributed by very reddened galaxies with either strong emission lines or an old stellar population component. The inset panel zooms over the primary redshift range of interest. The non-uniform distribution of the histogram highlights the fact that different SEDs are responsible for the signal in different redshift intervals.
 \label{fig:mszdist}}
\end{figure}

The resulting redshift distribution is shown in Fig. \ref{fig:mszdist}. 
The non-uniform distribution of the histogram is due to SEDs that satisfy the selection criterion only for portions of the redshift interval. For instance a 1 Gyr old solar metallicity model with modest extinction satisfies the criterion only for $z>5.5$. We identify three possible classes of low redshift interlopers at $z \sim0.6$, $\sim1.6$, and $\sim 3.2$. They are generally very reddened SEDs with either emission lines or an underlying old population. The total interloper fraction from these objects is found to be 23\% (17\% if one considered only $z<4$ as interlopers). This is only a rough estimate, as it does not account for the different galaxy populations having different and possibly evolving LFs. 
However, previous works show that the utilization of this library overpredicts the number of interlopers \citep[e.g. with respect to][]{malh05}; also, in the spectroscopic follow-up survey of the \vdrop galaxies in the Great Observatories Origin Deep Survey \citep[GOODS;][]{giav04a}, only $\sim 10\%$ of stars and low-$z$ galaxy interlopers were found \citep{vanz06}. Therefore, the real fraction of interlopers is possibly smaller than our conservative estimate. Note however, that we do not account for this source of systematic error in the rest of our analysis. 

\section{Catalogs Completeness and Simulations}
\label{sec:simulations}

The catalogs produced by SExtractor are subject to certain biases, which depend strongly on the chosen parameters. In addition to the detection settings, i.e., number of connected pixels and sigma-threshold above the background, the parameters that control the deblending and cleaning of sources are of particular importance. The correct interpretation of the LF that is obtained from the SExtractor-based catalogs requires a quantitative assessment of these biases. To this end, we performed two independent sets of simulations, the first based on idealized galaxy profiles and the second on the observed galaxies themselves. The simulations are described in detail in Appendix \ref{sec:detailssims}. Briefly:

\begin{enumerate}

\item In the first set of simulations (which we refer to as the "idealized-galaxies" simulations), we constructed an extensive set of exponential- and deVaucouleurs-profile galaxies, that were appropriately PSF-convolved and inserted at random positions into the NICP12 and HUDF images. SExtractor was then re-run with identical parameters as for the compilation of the original data catalogs, and the resulting object catalogs were used to identify detected, undetected or blended sources. This approach is similar in spirit to what was done by the GOODS team \citep{giav04a}.

\item In the second set of simulations (which we refer to as the "dimmed-galaxies" simulations), we generated a set of test images by co-adding to each original image a copy of itself, shifted by a few pixels and dimmed by a given amount. The composite images were rescaled to have the S/N vs. magnitude relation as the original image, and SExtractor was run on these final test images using the same parameters as for the extraction of the original data catalogs. This second set of simulations provides a complementary analysis to the above, as it offers the advantage of working with real rather than idealized galaxies, and does not depend on uncertainties in the adopted PSF. 

\end{enumerate}

\subsection{Input versus Output Properties}
\label{sec:invsout}
A problem in establishing the completeness of the source catalogs is that the true magnitudes of the sources are expected to be typically brighter than the measured ones, due to losses of flux from the wings of the galaxy light distribution. \citet{siri05} measured the flux losses for point-sources in the ACS camera filterbands and found that they strongly depend on the SED of the source, due to a stronger scattering halo of the PSF at longer wavelengths. Those authors find offsets of $\sim 0.1$ mag for $V_{606}$ and $i_{775}$, respectively, and $\sim 0.15$ mag for $z_{850}$ in apertures of radius $0.4\arcsec$. To take into account the extended nature of our catalog objects, we used our two sets of simulations to characterize the relation between input (true) parameters and output (measured) quantities. 

Idealized deVaucouleurs- and exponential-profile galaxies were generated with random $z_{850}$ magnitudes between 24 to 31. Two separate sets of tests were performed, the first using sizes\footnote{In all our analysis size always refers to half-light radius.} uniformly distributed in the range 0.05 to 0.5 arcseconds, and the second using sizes distributed according to a lognormal distribution centered at 0.25 arcseconds with $\sigma=0.3$. The results were essentially independent of the adopted input size distribution. In the following, we only use the simulations with uniformly distributed input sizes. Measuring the source magnitudes within an elliptical aperture of 2.5 Kron radii (SExtractor AUTO mag), compact sources were detected with 50\% probability down to 28.75 and 29 magnitudes in the NICP12 and in the HUDF images, respectively. As expected, the detected fluxes were typically smaller than the input ones. By integrating the theoretical deVaucouleurs and exponential profiles out to 2.5 times their Kron radii, the theoretically expected flux losses are found to be 9.6\% and 4.0\%, respectively, corresponding to 0.11 and 0.04 magnitude offsets. However, since the Kron radii measured by SExtractor are also underestimated, the offsets are expected to be even larger than those above. Indeed, in our simulations we found a magnitude-dependence for the amount of dimming, from $\sim 0.3$ mag and $\sim 0.07$ mag at $z_{850} = 24.25$ to $\sim 0.6$ and $\sim 0.2$ mag at $z_{850} = 28.25$ for the two kinds of profiles respectively. 

We also used the dimmed-galaxies simulations to obtain an independent estimate. In this case, the differences between dimmed and original magnitudes are expected to be smaller than in the case of idealized galaxies. The offset was indeed marginally smaller than for the exponential-profile idealized galaxies, i.e., in the range $\Delta \mathrm{mag} \simeq 0.04 - 0.12$. 

For very compact galaxies ($r_\mathrm{1/2} \lesssim 0\farcs15$), the measured sizes were generally overpredicted due to PSF blurring, while the opposite was true for larger objects. These biases showed a strong dependence on magnitude. The faintest idealized deVaucouleurs galaxies were typically measured to have $r_\mathrm{1/2} \sim 0\farcs1$ (i.e. to be unresolved at the resolution of the HST), independent of their original size. This offers a challenge for studying the size evolution of LBGs with redshift.

Furthermore, since we measure the colors by relying on the SExtractor dual-image mode, i.e. using the $z_{850}$ detection apertures, it is not adequate to apply a constant offset to the $V_{606}$ and $i_{775}$ magnitudes. The color measurements depend in fact on the $z_{850}$ flux as well as on the intrinsic color. Both the $i_{775} - z_{850}$ and the $V_{606} - i_{775}$ colors are affected in such a way that red sources have even redder, and blue sources even bluer, measured colors. This is not including the dependence of the PSF halo on the color.

In the light of these complex and uncertain corrections for dimming effects, we chose not to implement any correction to the measured catalog magnitudes. Thus, we note that our magnitude measurements might be underestimated by an amount, as estimated from our tests, of up to about 0.2 magnitudes. We found, however, that these light losses had only a marginal effect on our final LF when extending it with the SDF (see section \ref{sec:veff}); the resulting slope is steepened by about $\Delta\alpha\simeq0.03$.

\subsection{Detection Completeness}

There are two possible reasons why galaxies could be missing from our catalogs. First, a galaxy might not be detected due to its too low surface brightness; second, it could get blended together with another source and produce a false catalog entry. These two effects are strongly dependent on the SExtractor parameters.

\begin{figure}
\plotone{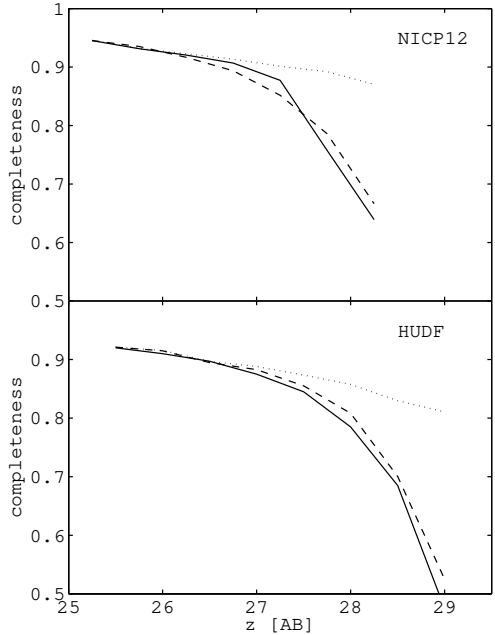}
\caption{Estimated detection completeness over the range of interest in the two fields as a function of observed magnitude from the idealized-galaxies (solid line) and dimmed-galaxies (dashed line) simulations. The completeness does not reach 100\% even at the brightest magnitudes, which is due to blending of sources by the SExtractor algorithm (dotted line, for idealized-galaxies). 
\label{fig:completeness_all}}
\end{figure}

We used both the idealized-galaxies and dimmed-galaxies simulations to estimate the strength of these effects in our catalogs and final samples of \vdrop galaxies. Specifically: (a) We associated idealized-galaxies with similar output (SExtractor) magnitudes and sizes to real galaxies randomly selected from our original catalog. We then estimated the completeness corrections for each magnitude bin by comparing the number of sources in each bin of SExtractor magnitude with the total number of idealized-galaxies of different input (i.e., theoretical) magnitudes and sizes that contributed to that specific bin of output magnitude. (b) Using the dimmed-galaxies simulations, we computed the magnitude-dependent completeness corrections by re-scaling the number of galaxies in each bin of output magnitude (i.e., recovered for the dimmed galaxies by SExtractor) to the number of sources of different input magnitudes (i.e., the theoretical values of the dimmed magnitudes) that contributed to that specific bin of output magnitude. Figure \ref{fig:completeness_all} shows the results of our tests. For both fields, the two independent estimates for the completeness in our samples, derived from the idealized- and dimmed-galaxies simulations, are in good agreement. At the magnitude limits of our selections, the NICP12 and HUDF catalogs are expected to be about 65\% and 50\% complete, respectively.

The probability that a given source is blended by SExtractor with another object depends on the magnitude of the source. 
In our simulation tests, when the SExtractor detection blended a simulated source with a real galaxy, we considered the simulated object as a detection when this was brighter than the original galaxy. Figure \ref{fig:completeness_all} also includes the incompleteness due to source blending; this is the reason why even our brightest sources are only complete at the 95\% level.

\section{The \vdrop Candidates}
\label{sec:vdropsources}

The application of the selection color window discussed above to our NICP12 and HUDF catalogs down to 28.5 and 29.25 magnitudes (with a ${\rm S/N} > 6$ cut) produces a total of 101 and 133 candidate $z\sim 5$ LBGs, respectively. Note that for sources whose \vFilter ISO fluxes were fainter than their 2$\sigma$-errors, we replaced the measured \vFilter fluxes with the corresponding 2$\sigma$ values in order to get a reliable lower limit on their \vi\ color.

To correct our catalog for both random and systematic photometric errors we adopted a Monte-Carlo technique. The amplitudes of the errors were estimated from both sets of simulations described in section \ref{sec:simulations}. The simulations show that systematic color shifts (i.e. input versus recovered color) are independent of the galaxy profile, unlike offsets in individual passband magnitudes. Therefore, before the application of random errors, the \vi and \iz colors of each catalog source were corrected for systematic errors; only a random error was instead applied to the \zFilter magnitude (see also section \ref{sec:invsout}). The selection of \vdrops was iteratively repeated and, after of order several hundred iterations, the mean number of \vdrop sources was computed in bins of \zFilter magnitude. Table \ref{tab:n_m} lists the mean number of \vdrop sources $N_{corr}$ per magnitude bin for both our fields which we use in the remainder of the analysis.

Tests were performed to ensure that the final LBG LF was independent of the adoption of the $2\sigma \ V_{606}$ flux limits for faint \vFilter sources. In particular, the calculations were repeated adopting $1\sigma$ and $3\sigma$ limits for the flux substitutions; no significant differences were found in the final LFs, as the probability of selecting a galaxy at any given redshift changed accordingly (see next section).

\begin{deluxetable}{cccccc}
\tablecaption{Number of detected LBGs and corrections in the two fields
\label{tab:n_m}}
\tablewidth{0 pt}
\tablecolumns{5}
\tablehead{\colhead{$m_{850}$\tablenotemark{a}} & \colhead{$N_{cat}$\tablenotemark{b}} & \colhead{$N_{corr}$\tablenotemark{c}} & \colhead{$C_\mathrm{Idealized}$\tablenotemark{d}} & \colhead{$C_\mathrm{Dimmed}$\tablenotemark{e}}}
\startdata

\cutinhead{NICP12}
25.25	&	6 & 4.0	 & 0.95 & 0.95 \\
25.75	&	4 &  5.3	 & 0.93 & 0.93 \\
26.25	&	12 &  10.8	 & 0.92 & 0.92 \\
26.75	&	14 &  12.9	 & 0.91 & 0.90 \\
27.25	&	12 &  17.8	 & 0.88 & 0.85 \\
27.75	&	29 &  21.3	 & 0.76 & 0.79 \\
28.25	&	24 &  23.0	 & 0.64 & 0.68 \\

\cutinhead{HUDF}
25.5 	&	3 & 3.0	& 0.92	&0.92 \\
26	&	2 & 4.0	& 0.91	&0.91 \\
26.5 	&	6 & 8.6	& 0.90	&0.89 \\
27 	&	23 & 17.3	& 0.87	&0.88 \\
27.5 	&	18 & 18.3	& 0.84	&0.86 \\
28 	&	24 & 23.0	& 0.78	&0.81 \\
28.5 	&	34 & 31.5	& 0.68	&0.70 \\
29 	&	23 & 24.8	& 0.47	&0.53 \\
\enddata
\tablenotetext{a} {central bin magnitude}
\tablenotetext{b} {uncorrected number of color-selected LBG sources}
\tablenotetext{c} {corrected number of \vdrop sources after Monte-Carlo resampling}
\tablenotetext{d} {completeness derived from the idealized galaxies simulations}
\tablenotetext{e} {completeness derived from the dimming simulations}
\end{deluxetable}

\section{The Luminosity Function of Lyman-Break Galaxies at $z\sim5$}
\label{sec:LF}

The rest-frame UV-continuum LF was derived from a maximum likelihood fit of a \citet{sche76} function to the observed number of \vdrop galaxies apparent magnitude bins, $N_i$:

\begin{equation}
N_i =  \int_{m_l}^{m_u} dm\ N(m; \phi_*, M_*, \alpha) =
\label{eq:LFintegral}
\end{equation}
\[
\int_0^\infty dz \frac{dV}{dz} \int_{m_l}^{m_u} dm\ p(m,z) \phi(M(m,z); \phi_*, M_*, \alpha) 
\]

Here $m_l$ and $m_u$ are the lower and upper bounds of the $i$-th magnitude bin, centred around $m_i$. The factor $p(m,z)$ is the probability that a galaxy with magnitude $m$ at redshift $z$ is detected and selected to be a $z\sim5$ LBG.

An "effective volume" ($V_{\mathrm{eff}}$) technique was also used to compute the stepwise LF in bins of absolute magnitude under the assumption that $M(m,z)$ is slowly-varying with redshift, i.e.: 
\begin{eqnarray}
N_i & \simeq & \phi(M(m_i,\overline{z})) \int_0^\infty dz \frac{dV}{dz} p(m_i,z) \\ 
 & \equiv & \phi(M(m_i,\overline{z})) \ V_\mathrm{eff}(m_i)
 \label{eq:veffLF}
\end{eqnarray}
Over the redshift range of interest, $z = 4.5 - 5.7$, the absolute magnitude at any given apparent brightness changes by about $0.5$ magnitudes, i.e., an amount comparable with the size of our magnitude bins. The resulting LF is thus generally less robust\footnote{Note that the situation would be exacerbated for an \iF-selected (rather than \zFilter-selected) \vdrop sample, since the differences in absolute magnitude over the same redshift range could be up to $\sim$2 magnitudes, i.e., much larger than the size of the bins, due to intergalactic \lya\ absorption.} than the one computed using Eq. \ref{eq:LFintegral}.

In order to properly constrain the value of $M_*$ in our best fit Schechter function, we adopt this version of our LF for matching it to the \citet{yosh06} data which constrained the $z\sim5$ LF at bright magnitudes in the Subaru Deep Field (SDF) (see Section \ref{sec:veff}).

The K-corrections required to transform the observed magnitudes into rest-frame 1400 \AA \  absolute magnitudes, $M(m,z)$, were computed assuming a 200 Myr old, continuously starforming galaxy template, reddened by $E(B-V) = 0.15$ with a \citet{calz00} dust extinction relation (hereafter referred to as "starburst" dust extinction). The differences in the K-corrections for alternative reasonable assumptions on the stellar population properties, including significantly younger stellar populations, were checked and found to be negligible ($\lesssim 0.05$ mag).

The probability that a galaxy with magnitude $m$ at redshift $z$ is detected and selected to be a LBG at $z\sim5$ can be decomposed into the product of two components, i.e., the completeness-probability $C(m)$ and the selection-probability $S(m,z)$, i.e. $p(m,z) \equiv C(m) \cdot S(m,z)$. We discuss this below.

\subsection{The Selection Probability $S(m,z)$}

The selection-probability as a function of magnitude and redshift was computed including the scatter in the photometry, which was derived from our simulations, and replacing with the $2-\sigma$ flux value the \vFilter magnitudes fainter than this limit. The input assumptions for the distribution of SEDs in the \vdrop population were based on the findings of \citet{stei99}. These authors inferred a distribution of spectral slopes from their sample of spectroscopically-confirmed $z\sim 3$ LBGs by assuming a 1 Gyr old, continuously starforming galaxy SED with varying dust content according to the starburst dust extinction relation \citep[see also][]{adel00}. The $E(B-V)$ distribution was found to be roughly approximated by a Gaussian with a median of 0.13 and a $\sigma = 0.12$. Note that this results in a small fraction of the galaxy population which is fitted by non-physical, negative values of $E(B-V)$; this might indicate the presence of strong \lya\ emission for some galaxies. Nevertheless, this "effective" distribution of $E(B-V)$ is in good agreement with the typical dust attenuation values found by \citet{papo01}, who used synthetic SED fitting to $z\sim 3$ LBGs to infer the properties of these galaxies. Since an age of 1 Gyr might be too old for $z\sim5$ galaxies, we used similar models with a 200 Myr old, constantly starforming SED \citep{bruz03}. The selection probability was found to be virtually independent of the assumed stellar age; this is understandable in the light of the dominant contribution from the youngest stellar populations to the UV emission.

\begin{figure}
\plotone{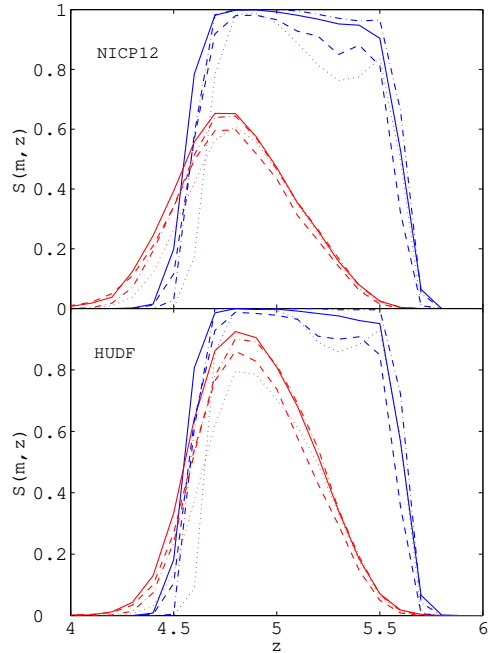}
\caption{\textit{Top $-$} The NICP12 selection probabilities for the different assumptions on the SED distributions and intergalactic hydrogen absorption discussed in the text, for the case of a bright ($25 - 25.5$, blue lines) and faint ($28-28.5$, red lines) magnitude bin. The solid (dot-dashed) and dashed (dotted) lines refer to the starburst (SMC) dust extinction relation, assuming the intergalactic hydrogen absorption prescription of Madau (1995) and Meiksin (2006), respectively. 
\textit{Bottom $-$ } The same for the HUDF.
\label{fig:smz_comparison}}
\end{figure}

To explore the effects of different assumptions for the intergalactic hydrogen absorption, the SEDs were attenuated according to the calculations of both \citet{mada95} and \citet{meik06}. In the latter case, the selection probabilities are slightly reduced, as a consequence of the smaller attenuation and thus smaller reddening of the SEDs. 

We estimated the impact on the selection probability of assuming that there are no Lyman-Limit systems along the line of sight attenuating the observed $V_{606}$ flux, a situation which is expected to happen with a probability of only $6\%$ up to $z\sim5$ \citep{meik06}. This effect was found to be negligible, i.e., the selection probability was reduced by less than $5\%$ around $z\sim5.2$. In the remainder of our analysis we therefore assume the estimated mean values for the intergalactic absorption.

We also tested alternative distributions for the UV slopes. \citet{verm07} used stellar population synthesis models to derive ages, masses, and dust attenuation properties of $z\sim5$ LBGs. Using the Small Magellanic Cloud (SMC) dust extinction law, both \citet{papo01} and \citet{verm07} find that the typical UV attenuation is smaller than in the case of the starburst extinction law. The exact values somewhat depend on the metallicities and Initial Mass Functions (IMFs). To estimate the selection probability, we adopted a 100 Myr old, continuously starforming template reddened by a Gaussian distribution of $E(B-V)$ with mean$=0.05$ and $\sigma = 0.04$, which is approximately in agreement with the distributions found by Papovich et al.\ and Verma and et al.\ Also in this case we explored the effects of changing the prescriptions for the hydrogen intergalactic absorption. Differences of order $10\%$ or less were found with respect to the previous estimates.

The selection probabilities for the NICP12 field, as derived under the various different assumptions, are shown in Fig. \ref{fig:smz_comparison} for a bright and a faint magnitude bin. In general, since the intergalactic absorption relation derived by \citet{meik06} is smaller than the corresponding calculation of Madau (1995), galaxies are less reddened assuming the former prescription; consequently, at any redshift of interest, they lie closer to the boundaries of our color selection. This leads to a reduced probability that they are selected as LBG candidates. The effect is exacerbated when the SMC extinction law is assumed: a considerable drop in the redshift distribution is expected around $z\sim 5.3$, and the lower limit for the selection-redshift is slightly increased. Note that the adoption of the $2-\sigma$ lower limit for the \vFilter fluxes leads to a considerably different redshift window for the fainter magnitude bin: the reddest, highest-redshift objects are not detected. Furthermore, due to the larger photometric scatter at fainter magnitudes, the redshift selection smears out to substantially lower redshifts, down to $z\sim4$. Note that we have not applied any correction for incompleteness in these plots; therefore, the drop in selection probability at fainter magnitudes that is observable in the Figure is exclusively due to the errors in the color measurements. Similar results hold for the HUDF, for which however the selection probabilities are somewhat larger, due to the smaller uncertainties in the color measurements (see bottom panel of Fig. \ref{fig:smz_comparison}). 

\subsection{The Impact of Different Physical Assumptions on the LF}
\label{sec:veff}

As we discuss below, different assumptions for physical input parameters such as the dust extinction law and the hydrogen intergalactic absorption relation have a non-negligible impact on key quantities such as the effective volume and thus the average luminosity density. It is thus of paramount importance that LFs derived from different data sets are first rescaled to identical assumptions for SED distributions, dust properties and intergalactic extinction, before they can be compared to study any possible evolutionary effect in the galaxy populations. In our analysis we adopt the Madau (1995) and starburst relations as our "fiducial" assumptions for the hydrogen absorption and dust extinction, respectively, as these are widely used in most previous works and allow for a direct comparison with the published results. We explore however below the effects on the LF parameters of varying this assumptions.

Our pencil-beam observations do not properly constrain the value of $M_*$ in the LF: given the rarity of very bright objects, the small volumes probed by the NICP12 field and the HUDF are not sufficient to adequately sample the knee of the Schechter function. We therefore rely on the SDF $z\sim5$ LF of Yoshida et al.\ (2006; see also Appendix \ref{sec:SDFyoshida}) to constrain the bright-end of our $V_{\mathrm{eff}}$ LF. These authors used similar fiducial assumptions for the intergalactic absorption and dust extinction relations, and similar techniques to correct for the observational biases, to those adopted in this paper.

\begin{figure}
\plotone{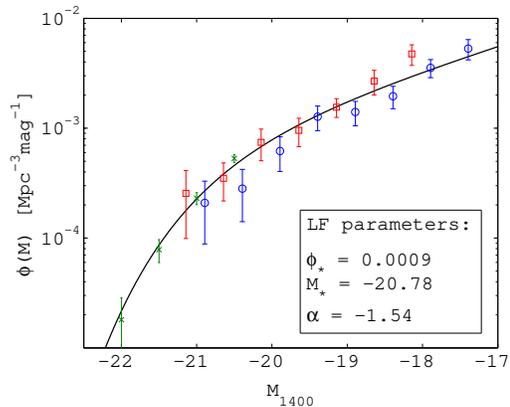}
\caption{Total $V_{\mathrm{eff}}$ LF obtained combining the NICP12 and HUDF data (this work) with the SDF LF \citep[$Ri'z'$-sample]{yosh06}. Red squares and blue circles correspond to the NICP12 and HUDF $V_\mathrm{eff}$ LFs respectively. Green crosses are the $V_\mathrm{eff}$-LF by \citet{yosh06}. The errorbars indicate Poissonian errors only. The solid line shows the best-fit Schechter function to the total LF (see text and Table \ref{tab:LF_others}). \label{fig:compositeLF}}
\end{figure}

The SDF LF is based on the $Ri'z'$ SUBARU filters, and is sensitive to a redshift window that is only slightly narrower than that of our data. It reaches down to $z'=26$ mag, i.e., $\sim 3$ magnitudes brighter than our measurements. In combining the SDF and NICP12/HUDF LFs, we neglected small K-corrections from 1500 \AA \ to 1400 \AA; the $V_{\mathrm{eff}}$ SDF LF nicely matches our $V_{\mathrm{eff}}$ NICP12 LF in the region of overlap. The total $V_{\mathrm{eff}}$ LF, that merges the SDF estimate with our NICP12 and HUDF LFs, spans over the 5-magnitudes range $M_{1400} = [-22.2, -17.1]$ (Figure \ref {fig:compositeLF}).

Adopting the completeness corrections derived from the idealized-galaxies simulations, the lowest $\chi^2$, best fit of a Schechter function to the total $V_{\mathrm{eff}}$ LF gives: $\phi_* = 0.9_{-0.3}^{+0.3}\cdot10^{-3}\ $Mpc$^{-3}$mag$^{-1}$, $M_* = -20.78^{+0.16}_{-0.16}$ mag, $\alpha = -1.54_{-0.10}^{+0.10}$ (see Figure \ref {fig:compositeLF} and Table \ref{tab:LF_others}). As expected, the value of $M_*$ is largely influenced by the SDF LF, while the slope $\alpha$ is constrained by our fainter data. The adoption of the completeness corrections derived from the dimmed-galaxies simulations, and of different SED distributions, dust extinction laws and intergalactic absorption prescriptions, has no significant impact on this result.

In particular, all tests performed by varying these quantities, independent of whether they include one or both of our two deep fields, return a value of $M_*$ that, within the errors, is consistent with $M_*=-20.7$. We therefore fixed $M_*$ to this value in order to highlight the effects of varying the input physical parameters on the LF computed according to Eq. (\ref{eq:LFintegral}) for the HUDF and NICP12 only.

\begin{deluxetable*}{ccccccc}
\tablecaption{$z\sim5$ Luminosity Functions \label{tab:LF_others}}
\tablewidth{0 pt}
\tablecolumns{7}
\tablehead{\colhead{Authors} & \colhead{Field} & \colhead{Area} & \colhead{$M_{\mathrm{lim}}$} & \colhead{$\phi_*$} & \colhead{$M_*$} & \colhead{$\alpha$}\\
& & \colhead{[arcmin$^2$]} &\colhead{[mag]} & \colhead{[Mpc$^{-3}$mag$^{-1}$]}&\colhead{[mag]} & }

\startdata

\citet{beck06} & HUDF & 11 & $-17.2$ & $1\cdot 10^{-3}$ & $-20.5$ & $-1.6$ (fixed) \\

\citet{yosh06}	&	SDF &  875	& $-20.3$ &	$0.58^{+1.04}_{-0.49}\cdot10^{-3}$	&	$-21.09^{+0.54}_{-0.74}$	& $-2.31^{+0.68}_{-0.60}$ \\

\citet{yosh06}	&	SDF &  875	& $-20.3$ &$1.23^{+0.44}_{-0.27}\cdot10^{-3}$	&	$-20.72^{+0.16}_{-0.14}$	& $ -1.82$ (fixed) \\[5pt]

This work\tablenotemark{a} + Yoshida & HUDF+NICP12+SDF & & & $0.9\pm{0.3}\cdot10^{-3}$ & $-20.78\pm0.21$ & $-1.54\pm0.10$ \\
This work\tablenotemark{b} + Yoshida & HUDF+NICP12+SDF & & & $1.0\pm0.3\cdot10^{-3}$ & $-20.76\pm0.18$ & $-1.62\pm0.08$ \\
\enddata

\tablenotetext{a}{Adopting the completeness correction derived from the idealized-galaxies simulations, the starburst dust extinction relation, and Madau (1995) intergalactic absorption prescription.}
\tablenotetext{b}{Adopting the completeness correction derived from the idealized-galaxies simulations, the SMC dust extinction law, and the Meiksin (2006) intergalactic absorption prescription.}
\end{deluxetable*}

The results obtained with fixing the value of $M_*=-20.7$ and excluding the SDF from the fits are listed in Table \ref{tab:diffLFs}. Only marginal differences are generally observed. Within the errors, the derived parameters are consistent with each other. Note however that the average luminosity density increases by about 20-30\% when the \vdrop population is described by the SMC dust extinction law and the Meiksin intergalactic absorption relation, relative to the more widely used combination of prescriptions which we have adopted as our fiducial model. This is due to a change of a similar order of magnitude in the resulting effective volumes. We note that the uncertainties in the completeness, estimated from our two complementary sets of simulations, have a comparably small effect on the faint-end slope of the LF as the different physical assumptions.

\begin{deluxetable*}{ccc|cc|cc}
\tablecaption{Impact of Physical Parameters on the LF (Eq. \ref{eq:LFintegral}; Constant $M_*=-20.7$)\tablenotemark{\dag} \label{tab:diffLFs}}
\tablewidth{0 pt}
\tablecolumns{6}
\tablehead{\colhead{Completeness} & \colhead{Dust Extinction} & \colhead{Intergalactic Absorption} & \colhead{$\phi_*$ NICP12} & \colhead{$\alpha$ NICP12}& \colhead{$\phi_*$ HUDF} & \colhead{$\alpha$ HUDF}}

\startdata
Idealized	& Starburst & Madau & $8.4 \cdot 10^{-4}$ & $-1.71$ & $5.8\cdot 10^{-4}$ & $-1.74$\\
Idealized	& Starburst & Meiksin & $9.5 \cdot 10^{-4}$ & $-1.71$& $6.4\cdot 10^{-4}$ & $-1.75 $\\
Idealized	& SMC & Madau & $9.0 \cdot 10^{-4}$ &$-1.71$ & $6.0\cdot 10^{-4}$ & $-1.75$\\
Idealized	& SMC & Meiksin & $11.6 \cdot 10^{-4}$ & $-1.66$& $7.0\cdot 10^{-4}$ & $-1.76$ \\[4pt]

Dimmed	& Starburst & Madau & $8.9 \cdot 10^{-4}$ & $-1.68$ & $6.0\cdot 10^{-4}$ & $-1.71$\\
Dimmed	& Starburst& Meiksin & $10.0 \cdot 10^{-4}$ & $-1.67$ & $6.5\cdot 10^{-4}$ & $-1.72$\\
Dimmed	& SMC & Madau & $9.4 \cdot 10^{-4}$ & $-1.68$ & $6.1\cdot 10^{-4}$ & $-1.72$\\
Dimmed	& SMC & Meiksin & $12.1 \cdot 10^{-4}$ & $-1.63$ & $7.1\cdot 10^{-4}$ & $-1.73$

\enddata
\tablenotetext{$\dagger$}{Typical errorbars for $\phi_*$ and $\alpha$ are $\pm 2.1 \cdot 10^{-4}$ and $\pm 0.16$, respectively. The starburst dust model assumes $E(B-V) = 0.13 \pm 0.12$; the SMC dust model assumes $E(B-V) = 0.05 \pm 0.04$.}

\end{deluxetable*}

\section{The Role of Cosmic Variance}
\label{sec:cosmicscatter}

\subsection{The Underdensity of the HUDF}

On average, independent of the assumed input parameters, the LF derived from the HUDF results in a lower average luminosity density: The HUDF appears to be "under-dense" with respect to the NICP12 field (see also Fig. \ref{fig:compositeLF}). By integrating the LF from Table \ref{tab:diffLFs} over the magnitude range of overlap between the HUDF and the NICP12 field, i.e., $M_{1400}=[-21, -18]$, the amount of this underdensity is about 30\%, independent of the specific input physical assumptions. Although a larger number of low-redshift interlopers in the NICP12 sample cannot be ruled out, (a) the homogeneity in the procedures with which both data sets have been handled and analyzed, and (b) theoretical expectations concerning the amount of scatter amongst volumes of order of those sampled with the HUDF and NICP12 field, both support the interpretation that the under-density in $z\sim5$ LBG of the HUDF has a physical origin. Note that, in their analysis of the $i$-dropouts population, \citet{bouw05} find the HUDF under-dense at the bright end, also by about $\sim30\%$, relative to the GOODS field. Thus, it appears that the $\sim30\%$ under-density of the HUDF extends at least over the volume encompassing the entire $z=4.5 - 6.5$ epoch.

\subsection{Quantification of Cosmic Variance Effects in Our Study}

In order to estimate the impact of cosmic variance on our measurements, we use a $512^3$-particles, dark-matter-only cosmological simulation in a box of $100 Mpc/h$ side, that results in a particle mass of $5 \cdot 10^8 M_{\sun}/h$. Simple prescriptions to populate the dark matter halos with luminous galaxies, and a realistic beam-tracing algorithm were used to study the field-to-field variations in the number counts of star forming galaxies at the epoch of interest.

The numerical simulation was carried out with the public
 PM-Tree code Gadget-2 \citep{spr05} adopting a 
 $\sigma_8=0.75$ value. The initial conditions were generated using a code based on the Grafic algorithm
 \citep{bert01} with a $\Lambda CDM$ transfer function
computed according to the fit of \citet{eis99}, and assuming a spectral index $n=1$. 
Dark matter halos were identified in the simulation snapshots using the HOP halo finder \citep{eis98}; the redshift interval $4.5-6$ was sampled in steps of $\Delta z =0.125$. 
The simulation reproduces a volume about 73 times larger than the effective
volume probed by each of our two fields for the \vdrop population ($\approx 5.7 \times 5.7
\times 420 (Mpc/h)^3$).

Source catalogs were constructed for each snapshot, and analyzed with our beam-tracer algorithm, which is similar to the
one described in \citet{kit06}. We present details of the beam-tracer algorithm in
\citet{tre07}. In brief, the pencil beam associated with the ACS field
of view is traced through the simulation box across different snapshots
of increasing redshift. As the uncertainty in the
\vdrop redshifts is larger than the size of the simulation box,
the simulation box needs to be beam-traced several times. We carried out $4000$ Monte
Carlo realizations of the beam-tracing procedure by varying the initial
position of the beam; the beam was always angled so as to guarantee
no overlap between different beam-tracing runs. 
With our choice of angles we got a separation between different beam regions of at least $\sim$15 Mpc/h, which corresponds to $\sim$540 arcsec at $z=5$. We verified that fields with such separations are essentially statistically independent. Note also that our 100Mpc/h box is different from an ideal simulation of a 420Mpc/h box as in the power spectrum we are missing contributions from wavelengths greater than 100Mpc/h. For the scales we are interested in at $z>4$, the missing power is however negligible in the two-point correlation function and hence in the cosmic variance.

Dark matter halos within a beam were flagged, and a simple
model was used to determine the number of \vdrops within a flagged
halo. In particular, the intrinsic number of galaxies was computed using the halo
occupation model description. This gives an average occupation number
\citep{wec01}:

\begin{equation}
\langle N_{gal} \rangle = \theta(M-M_{min})(1 + (M/M_1)^{\alpha}),
\end{equation}

where $\theta(x)$ is the Heaviside step function, $M_{min}$ is a
minimum halo mass threshold, $M_1$ is the typical scale where multiple
galaxies are present within the same halo, and $\alpha>0$ is of 
order unity. If $M>M_{min}$, in addition to the one galaxy at the center of a
halo, a suitable number of companion galaxies is extracted from a Poisson
distribution with mean $(M/M_1)^{\alpha}$. A
fraction $\eta$ of the total galaxies in each halo is finally identified as \vdrops, so as to take
into account that \vdrops may be detected for only a limited period during
their evolution.
In addition to this, we tossed 50\% of the selected galaxies due to detection and selection incompleteness present in our survey.

\begin{figure}
\epsscale{0.75}
\plotone{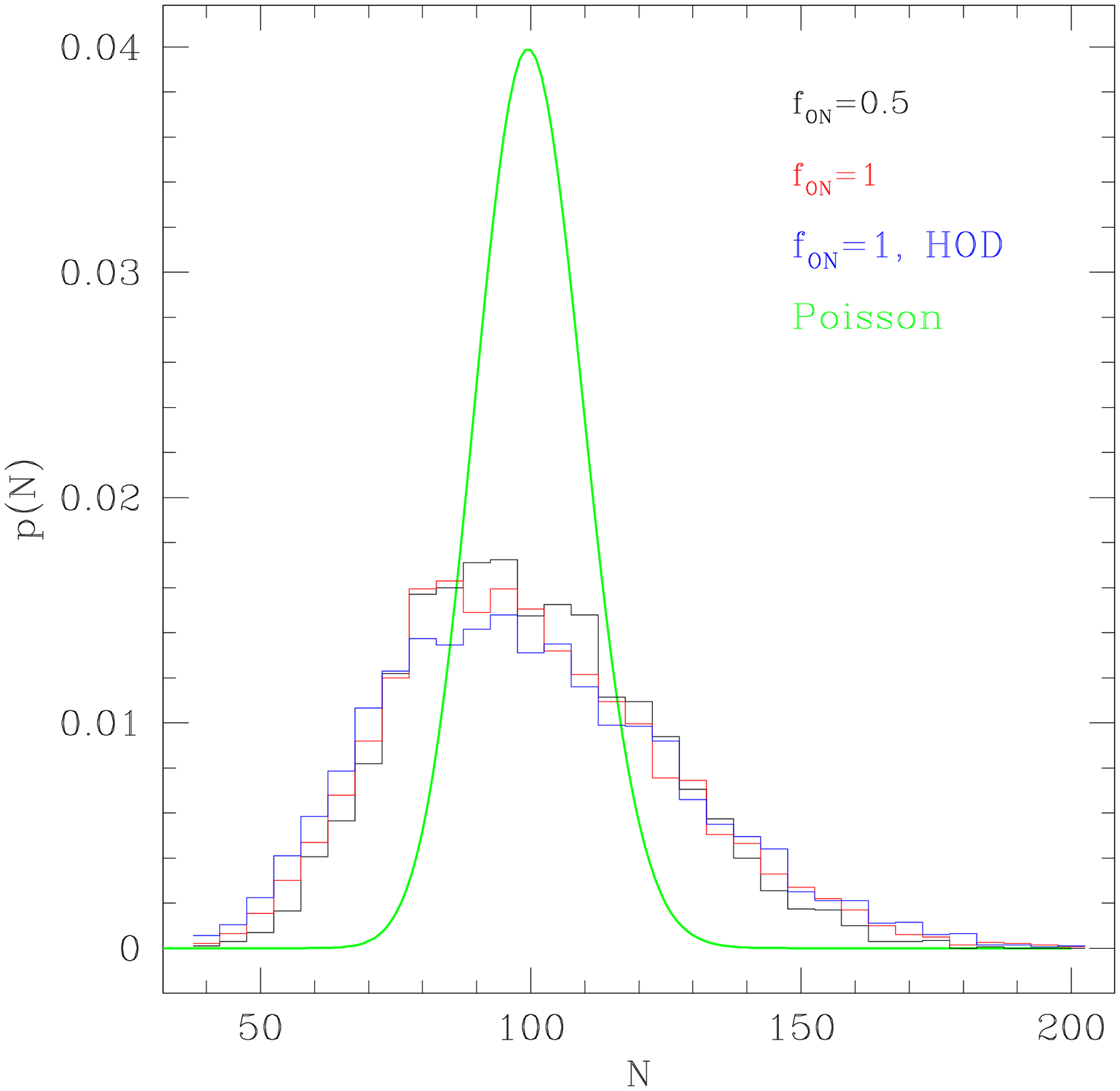}
\plotone{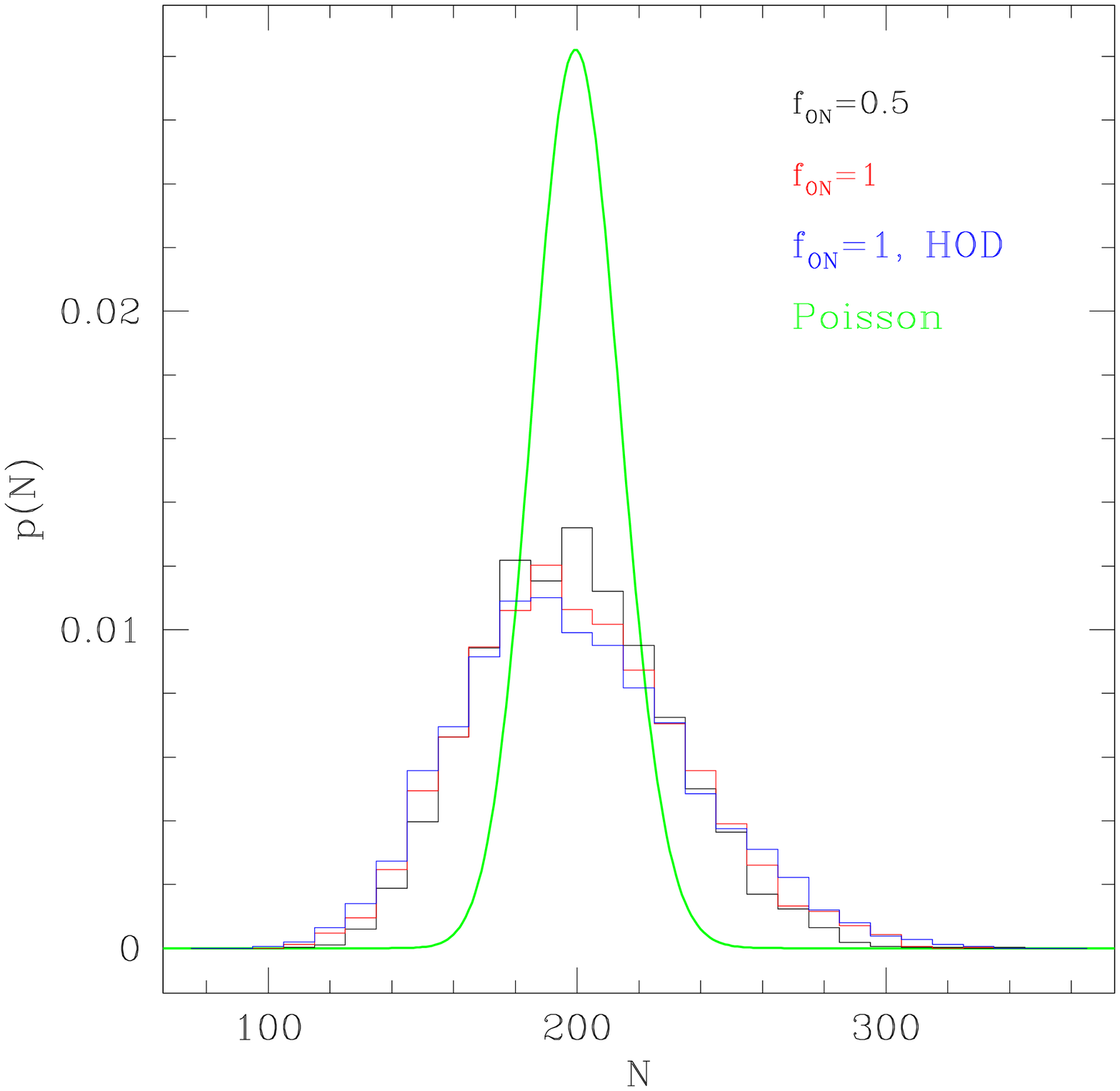}
\caption{Probability distribution of the \vdrops number counts in
one ACS field (top) and in the two HUDF + NICPIC12 fields (bottom),
accordingly to different halo occupation distribution models. Black line: 
one galaxy per halo, $\eta=0.5$; red line: one galaxy per halo
$\eta=1$; blue line: HOD model with $M_1=5\cdot 10^{11}
M_{\sun}/h$ and $\eta=1$. The Poisson distributions are overplotted in green for comparison. }\label{fig:scatter}
\end{figure}

The results for different choices of the model parameters are reported
in Fig.~\ref{fig:scatter}. We consider: (i) $\eta=1$, $M_1 \to
+\infty$ (i.e., one galaxy per halo); (ii) $\eta=0.5$, $M_1 \to
+\infty$ (i.e., 50\% probability that a DM halo hosts a LBG galaxy);
(iii) $\eta=1$, $M_1 = 5\cdot 10^{11} M_{\sun}/h$ and $\alpha=1$. The
value of $M_{min}$ is kept as a free parameter and it is adjusted to
have the same average number of counts in all three cases. Variations
of $M_{min}$ are limited in the range $M_{min} = 2- 3.5 \cdot 10^{10}
M_{\sun}/h$.

The distribution of the number counts probability is very
similar in the three cases, and the field-to-field standard deviation
is $\sigma = 25 \pm 2 \%$. This is significantly larger than the Poisson
uncertainty of $10 \%$ for a value of $\sim100$ counts per field, which is about the number of \vdrop counts/field that we have detected. 
Depending on the specific model, there are minor differences in the
computed standard deviation ($\Delta \sigma \approx \pm 2\%$): for
$\eta < 1$, a smaller $M_{\min}$ is required; therefore, the variance
decreases, as lower-mass halos become less clustered. As expected, the
largest variance results in the case of multiple galaxy occupancy of the halos.

We also computed the relative variations in expected counts for two fields which
are separated by about $\approx 10$ arcmin, as is the case for the HUDF and the NICP12 field. 
We find that the two fields have essentially independent counts, i.e., the
variance of the average number of detected \vdrops decreases by a
factor $\sqrt{2}$. 

Finally it is interesting to highlight that the
narrow-beam geometry of our data significantly reduces the cosmic
scatter with respect to a similarly-sized, spherical volume for a single field, which has a variance in the number counts of $\approx 50\%$ (see also \citealt{some04}). This is due to the stretched redshift
space of the pencil beam, that probes a variety of environments by minimizing the probability of enclosing only a large void or a rich cluster.

\section{Discussion}
\label{sec:discussion}

We discuss below the implications of our measurement of the faint-end of the $z\sim5$ LF of star forming galaxies for the evolution of galaxies, in the context of current models of galaxy formation.

\subsection{Evolution of the Faint-End of the LBGs Luminosity Function between $z\sim3$ to 6}

\begin{figure}[htb]
\plotone{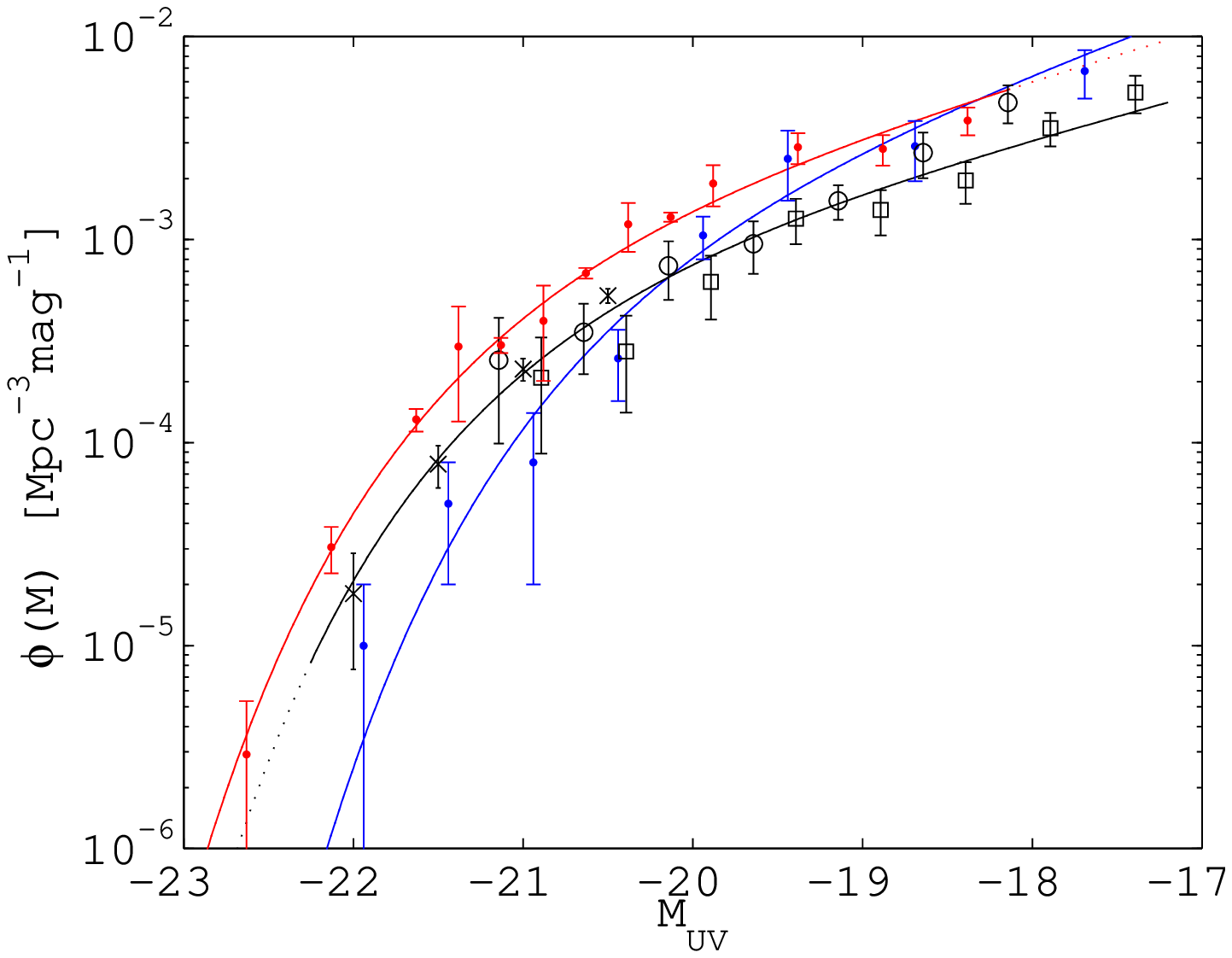}
\plotone{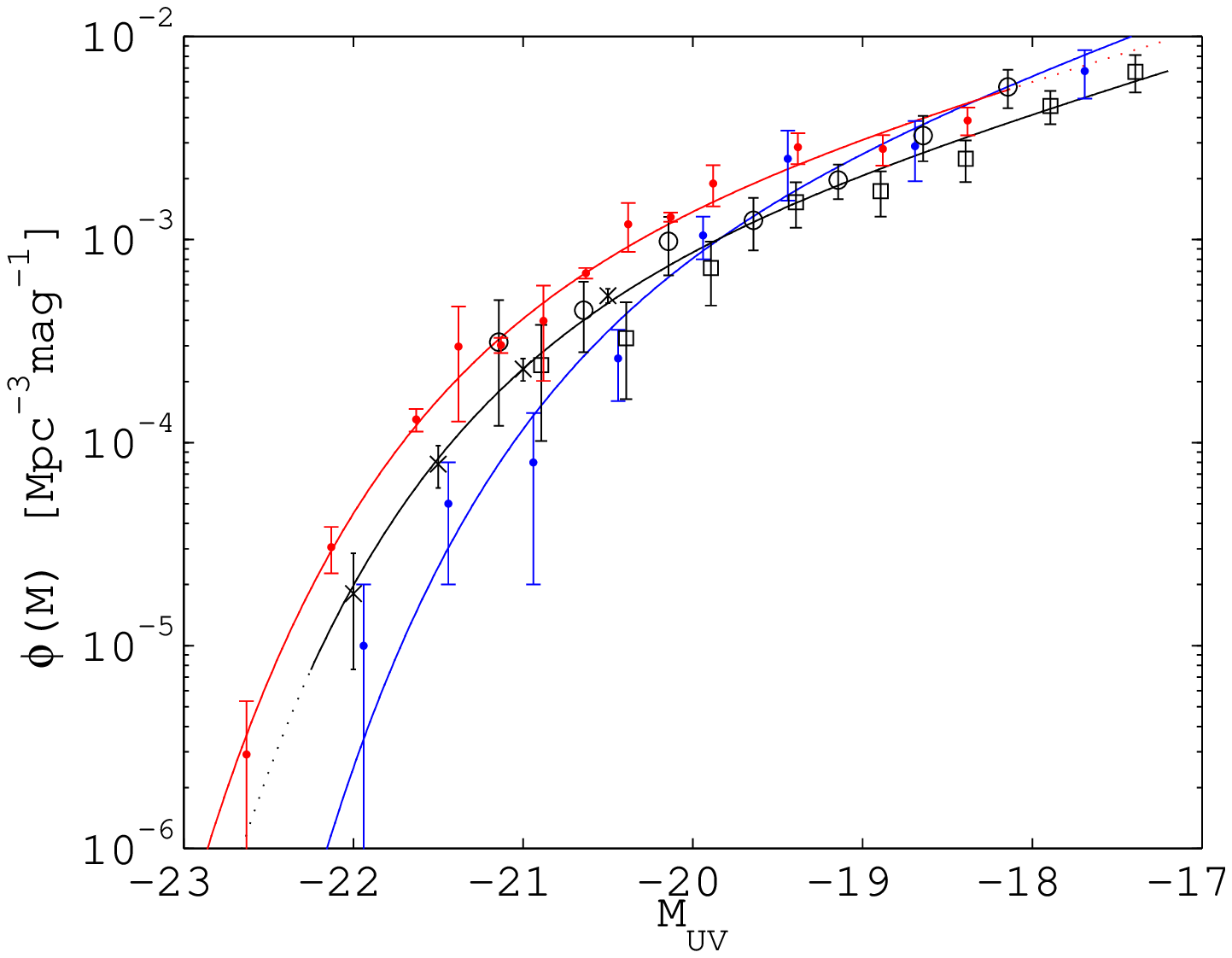}
\caption{Evolution of the rest-frame 1400 \AA\ LBG LF in the redshift range $z\sim3-6$. 
\textit{Top panel:} The LFs assume the starburst dust extinction relation and the intergalactic absorption prescription of Madau (1995). Our fiducial $z\sim5$ LF is represented by the black circles (NICP12 data) and the black squares (HUDF data), plus the black crosses, which represent the $z\sim5$ SDF LF of Yoshida et al. (2006). In blue we show the
 $z\sim 6$ LF of \citep{bouw05}, and in 
red the $z\sim3$ LF of \citep{stei99}. Solid and dotted lines are used to indicate the magnitude baseline over which the LF were measured and extrapolated, respectively. 
\textit{Bottom panel:} A similar plot as above, in the assumption of a SMC dust extinction law and a Meiksin (2006) intergalactic absorption correction. The parameters for the LFs are listed in Table \ref{tab:LF_others}.
\label{fig:LFevol}}
\end{figure}

There have been several other studies of the LBG LF from $z=3-6$. While for $z\sim3$ the different LFs and number counts seem to be in reasonable agreement with each other \citep[e.g.][]{stei99,sawi06,hild07}, at $z\sim6$ largely different LFs were reported in the literature \citep[e.g.][and see also discussion in Bouwens et al. 2005]{Bunker04,dick04,Windhorst04,malh05}. This might be mainly due to the highly model dependent corrections needed for the computation of the $i$-dropout LF. In order to investigate the evolution of the LBG LF over cosmic time, we use the analyses of \citet{stei99} and \citet{bouw05} at redshift $z=3$ and $z=6$, respectively, since these are widely used for comparisons in other works, and represent an ideal compromise between area coverage and depth. These are plotted in Figure \ref{fig:LFevol}. The $z\sim3$ LF is K-corrected to 1400 \AA\ and adapted to our cosmology ($M_*(z=3)=-20.9$); it is based on identical assumptions as in our fiducial LF, with which it can thus be directly compared. The $z\sim6$ LF is based on a "cloning" algorithm, which is well bracketed by our two different assumptions for the dust extinction and the hydrogen intergalactic absorption.

The possible underdensity of the HUDF, and the associated uncertainty in the faint-end slope of the $z\sim5$ LF, is relevant in establishing the evolution of such faint-end through the comparison with the $z\sim3-6$ estimates. In particular, the upper panel of Figure \ref{fig:LFevol} shows that the inclusion of the HUDF in the computation of the $z\sim5$ LBGs LF leads to a shallower faint-end slope than at $z\sim3$; in contrast, the faint-end slope that we derive using the NICP12 field alone for the $z\sim5$ LBGs is very similar to the one measured at that later epoch. A similar result is found when the SMC extinction law and the Meiksin intergalactic extinction relation are assumed, as shown in the lower panel of the figure. This result is thus robust also towards possible variations of the dust extinction properties through cosmic time. The possibility remains that the HUDF is, as we discuss above,  $\sim 30\%$ under-dense relative to the NICP12 (and the mean); a renormalization of the HUDF by such an amount would actually lead to a slightly steeper faint-end slope at $z\sim5$ than at $z\sim3$.

Rather similar conclusions are drawn for the redshift evolution of the faint-end of the LF in the $z\sim5$ to $z\sim6$ redshift window. Inspecting both panels of Figure \ref{fig:LFevol}, our fiducial $z\sim5$ faint-end slope is consistently shallower than the $z\sim6$ estimate, for both the HUDF and the NICP12 field; the NICP12 field provides however, at $z\sim5$, a similar faint-end slope to the one measured at $z\sim6$ when the SMC dust extinction law and the Meiksin intergalactic absorption correction are adopted. A similar agreement between the $z\sim5$ and $z\sim6$ faint-end slopes would also be obtained if the HUDF were renormalized to account for a $\sim 30\%$ underdensity caused by cosmic variance. Note that \citet{bouw05} did renormalize the $z\sim6$ HUDF number density, increasing it by about 32\%, based on the comparison with the GOODS data.

\subsection{Comparison with Theoretical Predictions}

We compare our observed $z\sim5$ LBG LF with theoretical predictions from
the GALFORM semi-analytical model \citep{cole00}, which calculates
galaxy formation in the framework of the CDM model of structure
formation. It includes the assembly of dark matter halos by mergers,
heating and cooling of gas in halos, star formation from cold gas,
feedback from supernovae and photoionization, galaxy mergers, and
chemical evolution of stars and gas. Star formation occurs in two
modes: quiescent in disks, and starbursts triggered by galaxy
mergers. Galaxy luminosities are calculated based on stellar
population synthesis, and include obscuration by dust, which is
calculated self-consistently by radiative transfer using the GRASIL
model \citep{Silva98,Granato00}. The version of the model which we
use is that described in \citet{baug05} \citep[see also][]{lacey07a},
and is based on the $\Lambda$-CDM cosmology, with $\Omega_0= 0.3$,
$h=H_0/(100 km s^{-1} Mpc) = 0.7$, and $\sigma_8= 0.9$. An important
difference of this model from the earlier \citet{cole00} model is the
much larger role of star formation bursts at high redshift. In the
\citet{baug05} model, bursts are responsible for 30\% of the total
star formation when integrated over all redshifts, compared with
around 5\% in the \citeauthor{cole00} model, and bursts dominate the
total star formation rate density at $z>3$. This difference in
behaviour results from (i) modifying the quiescent star formation
time-scale in galactic discs, in order to make mergers at high
redshift more gas-rich; and (ii) allowing triggering of bursts by both
minor and major mergers. The predictions of this model for LBGs at
different redshifts will be presented in full in \citet{lacey07b},
here we give only the results for $z\sim5$. The predictions of this same
model for Ly$\alpha$-emitting galaxies at similar redshifts have
already been given by \citet{LeD06}.

The Initial Mass Function (IMF) has a large impact on the predicted
$z\sim5$ LF. We therefore computed the results for two different
assumptions, namely (1) a model where {\it all} stars are produced
with a \citet{kenn83} IMF, and (2) a model where quiescent star
formation is described by a Kennicutt IMF, but star formation in the
merger-induced starbursts occurs with a "Top-Heavy" ($x=0$) IMF. The
Kennicutt IMF has $x=0.4$ for $m<M_\odot$ and $x=1.5$ above this
threshold; it is therefore quite similar to a \citet{salp55} IMF for
$m>M_\odot$. An IMF similar to the Kennicutt IMF has been argued to better fit the
observational data from our Galaxy and other spiral galaxies than a
Salpeter IMF \citep[e.g.][]{Scalo98}. At constant stellar mass, the
far-UV (1000-3000\AA) luminosity for a Kennicutt IMF is within 10-20\%
of that for a Salpeter IMF. The primary motivation in \citet{baug05}
for introducing a top-heavy IMF in starbursts was in order to explain
the number of sub-mm galaxies seen at high redshift - with a normal
IMF, the predicted number counts of faint sub-mm galaxies were too low
by more than an order of magnitude. However, a top-heavy IMF in
starbursts is also motivated by theoretical calculations
\citep[e.g.][]{lars98}, which suggest that, under the conditions
that exist in starbursts, the IMF could be biased towards more massive
stars relative to quiescent star formation.

\begin{figure}[htb]
\plotone{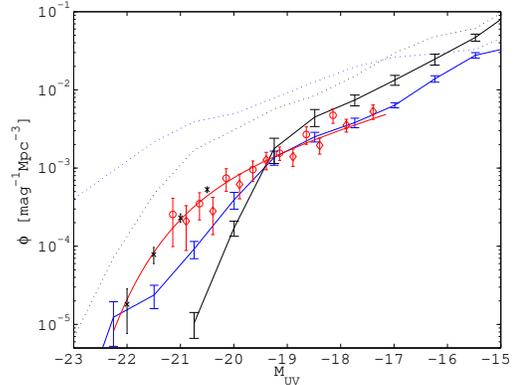}
  \caption{Comparison of our fiducial $z\sim5$ LF (red solid line: fit
to the NICP12, HUDF and SDF data; red circles: NICP12 data; red
diamonds: HUDF data; black crosses: SDF data) with the prediction of
Lacey et al.\ (2007). The simple-Kennicutt and Kennicutt+Top-Heavy
IMF models are shown in black and blue, respectively; dotted and solid
lines indicate unobscured and dust-extincted models. K-corrections from 
1500\AA\ to 1400\AA\ are negligible and were not adopted here.} \label{fig:1}
\end{figure}

Figure \ref{fig:1} shows the rest-frame 1500\AA\ LF at $z\sim5$ predicted
by the semi-analytical models \citep{lacey07b}, compared to our
measurements. 
The error bars on the theoretical curves show bootstrap estimates of
the uncertainty due to the finite number of galaxies in the simulated
samples; the numerical runs are very CPU intensive, and it is thus
currently impractical to greatly reduce these error bars by increasing
the sample size. The dotted black and blue lines in the figure show,
for the simple-Kennicutt and Kennicutt+Top-Heavy IMFs, respectively,
the predicted LFs without dust obscuration. These substantially
overpredict the number density of $z\sim5$ LBGs over the whole luminosity
range sampled by the data, but this is not surprising, since the
average dust extinctions of star-forming galaxies in the far-UV at
high redshift are thought to be quite large.

The solid black and blue lines are the simple-Kennicutt and
Kennicutt+Top-Heavy IMFs models including the effects of dust extinction. 
At the bright-end of the $z\sim5$ LF, the simple-Kennicutt
model with dust extinction substantially underestimate the number
density of bright ($M_{UV}<-19$) $z\sim5$ LBGs. The Kennicutt+Top-Heavy
IMF model is much closer to the observational data. Although this
latter model still appears somewhat low compared to the observations,
this could be explained by relatively small changes in the dust
extinction, to which the model predictions are quite
sensitive. Resolving this issue may require better observational
estimates of the dust extinctions in LBGs. 
Another possibility would be that either the fraction of mass produced
in bursts or the fraction of high mass
stars in the bursts is even higher than in our Top-Heavy IMF model.

The Kennicutt+Top-Heavy IMF model also appears to give a somewhat better
fit to the faint end of the observed $z\sim5$ LF than the
simple-Kennicutt model, although the difference between the two models
at the faint end is only slightly larger than the scatter in our
estimate due to completeness corrections, input physical parameters,
and possibly cosmic variance.

\section{Summary and Concluding Remarks}
\label{sec:conclusion}
We have presented the UDF05 project, a follow-up of the HUDF aimed at increasing the area of the HUDF and securing two more fields with ultra deep ACS and NICMOS imaging. 

In this first paper we use the UDF05 to estimate the $z\sim5$ Luminosity Function of Lyman-Break Galaxies down to very faint absolute magnitudes. Specifically, we use the NICP12 ACS field of the UDF05, and the ACS HUDF for a comparison, to extend the measurement of the LF of star forming galaxies at $z\sim5$ down to $M_{1400}=-17.1$. This allows us to constrain the faint-end slope of the $z\sim5$ LF very accurately to $\alpha \sim -1.6$, which is in good agreement with the results of Giavalisco et al. (private communication) based on the GOODS data to a similar depth as in our work. After this paper was submitted, \citet{bouw07} published LBG LFs from $z\sim4-6$ based on all the available deep ACS fields and also their results are in very good agreement with our $z\sim5$ LF. 

Attention was paid to correct the raw number counts for selection biases and photometric errors, and to estimate the impact on the LF of different assumptions for the properties of the $z\sim5$ LBG population. 
We find that different assumptions about the LBG galaxy population and intergalactic absorption result in an uncertainty in the density of LBGs at $z\sim5$ of about 25\%. 

Under similar physical assumptions, the HUDF is underdense in $z\sim5$ LBGs by about 30\% with respect to the NICP12 field, a variation which can be accounted for by the expected amount of cosmic variance in the pencil-beam volumes we have probed. 

A substantial steepening of the faint-end slope is observed from the local Universe out to $z=3$ \citep{ryan07}, in agreement with semi-analytical predictions within a hierarchical galaxy formation scenario \citep{khoc07}. Further steepening of the faint-end slope above $z\sim 3$ has been reported by \citet{iwat07}. As discussed by \citet{ouch04}, however, the selection window of \citet{iwat07} is rather broad, and it is thus likely that their sample contains a large number of low-redshift interlopers (see also Appendix \ref{sec:SDFyoshida}). This would lead to an overestimation of the bright-end of the LF and thus to a spurious evolution of the LF. Our much deeper data, extending about 3 magnitudes fainter than the \citet{iwat07} sample, show in contrast no evolution of the LF faint-end of LBGs in the $z\sim3$ to 6 redshift window, and substantiate a picture in which, once cosmic variance is taken into account, the faint-end of the LF of LBGs remains constant throughout this cosmic period.  This is particularly interesting in comparison with the significant evolution shown, over the same redshift range, by  the bright Lyman-break galaxy population \citep[see also e.g.][]{yosh06,bouw07}.

A comparison with our semi-analytical models, specifically aimed at investigating the effect of the IMF, suggests the possibility that the IMF of $z\sim5$ LBGs is more top-heavy than a Salpeter IMF. Other factors could, however, be relevant, such as variations in the dust extinction corrections.

\acknowledgments
Acknowledgements: We wish to thank the referee for helpful comments that have improved the presentation of our results. This work has been partially supported by NASA HST grant 01168.

Facilities: \facility{HST(ACS)}.

\appendix

\section{Comparison with Published $z\sim5$ LBGs Luminosity Functions}
\label{sec:comparison}

\subsection{The Subaru Deep Field}
\label{sec:SDFyoshida}
Ground-based Subaru observations have been used to derive the $z\sim5$ LF of LBGs \citep{iwat03,ouch04,yosh06}. In particular, \citet{yosh06} extended the analysis of \citet{ouch04} on the SDF with observations that reached 0.5 mag fainter and were wider in area by a factor of 1.5. As both authors adopted very similar techniques and their number counts of $z\sim5$ galaxies are in very good agreement, we focus on the results by \citet{yosh06} only. Using the combination of Subaru $V,\ R,\ i',\ z'$ filters, these authors defined two samples of $z\sim5$ LBGs, whose reliability was tested with spectroscopic data. We can directly compare with their $Ri'z'$ analysis, as this filter combination is sensitive to a very similar redshift window ($z=4.5-5.4$) as our data.

Following a procedure similar to the one adopted in this paper, \citet{yosh06} estimated the selection probability and completeness by adding artificial galaxies to the original images; they did not correct their samples for photometric errors as we do, but estimated the number of expected low-$z$ interlopers by simulating galaxies with colors according to a photo-$z$ catalog from the Hubble Deep Field North \citep{will96}. 

The best-fitting LF derived by \citet{yosh06} is listed in Table \ref{tab:LF_others} (only $Vi'z'$ sample, for the cases of keeping all LF parameters free and of fixing the faint-end slope to their $z\sim4$ result). In both cases the faint-end slopes are very steep, i.e., significantly steeper than the one that we measure from our 3-magnitudes deeper data. This highlights the need for very deep imaging surveys to study the contribution to galaxy assembly at high redshifts by galactic systems well below the few bright percent of the star forming population at those epochs.

Note that we did not use the LF of \citet{iwat03} which has a larger amplitude due to a possibly significant contamination by low redshift interlopers \citep[see also][]{ouch04,capa04}.

\subsection{A Previous Analysis of the Hubble Ultra Deep Field}

\citet{beck06} present the first attempt to derive the $B_{435}$- to $i_{775}$-dropout LFs in the HUDF. Their approach is very different from ours, as they did not correct for observational biases on the selection of dropout sources. Instead, they assumed that these biases affect all dropout samples equally, and base on this assumption their study of the evolution of the LBGs LF in the $z\sim4-7$ redshift window. They extract an $i_{775}$-selected sample from the GOODS v1 data to constrain the bright-end of the LF, and derive the following best fit parameters: $\phi_* = 0.001$ Mpc$^{-3}$mag$^{-1},\ M_*(1400 $ \AA$) = -20.5,\ \alpha = -1.6$ (fixed). 

\begin{figure}[htbp]
\epsscale{0.5}
\plotone{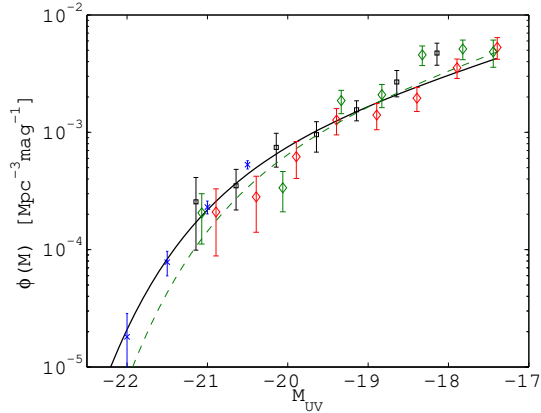}
\caption{Comparison of our LF to the published HUDF LF by \citet{beck06}. Our best fit to the combination of the HUDF data (red diamonds), NICP12 data (black squares), and the SDF data (blue crosses) is shown as a black line, and the estimates of Beckwith et al.\ are shown in green (symbols and line for the measurements and best fit Schechter function, respectively). 
\label{fig:LFcompare}}
\end{figure}

Figure \ref{fig:LFcompare} shows our fiducial $z\sim5$ LF in comparison with the estimate of Beckwith et al. The two LFs are in good agreement, except for one datapoint at $M=-18.2$.  This is a very interesting result, as it suggests that the several effects, which we have corrected for in our analysis, likely cancel out, thereby validating the \citet{beck06} assumption, at least for the $z\sim5$ LBG population. In a future paper (Oesch et al.\ 2007, in preparation) we discuss whether a similar conclusion can be drawn at $z\sim6$.

\section{Simulations: Details}
\label{sec:detailssims}

\subsection{Idealized-Galaxies Simulations}
\label{sec:simdetails}

Two different types of light profiles were considered for the ideal-galaxies simulations,
namely an exponential and a deVaucouleurs profile. These distributions are expected to more or less bracket the whole range of galactic light profiles. Specifically, the surface brightness distribution of disk galaxies was approximated by:

\begin{equation}
I(x,y)=\frac{L_{tot}}{2\pi \,a\,b} e^{-\sqrt{x^2/a^2+y^2/b^2}} 
\end{equation} 
where $a$ and $b$ are the major and minor axes lengths, the inclination angle $i$ is given by $\cos(i)=\frac{b}{a}$, and $L_{tot}$ is the extrapolated total galaxy light. The coordinates $x$ and $y$ are measured from the centre of the galaxy. The disk was assumed to be infinitely thin, which would lead to unrealistic shapes for edge-on systems. To avoid this unphysical situation, galaxy shapes were simulated with inclinations up to a maximum of $80^\circ$. To retain a flat distribution of inclinations in the entire 0$^\circ$ to 90$^\circ$ range, inclinations $> 80^\circ$, drawn from an assumed flat distribution, were set equal to 80$^\circ$.

The deVaucouleurs profile adopted to simulate spheroidal-type galaxies reads:

\begin{equation}
I(r)=L_{tot}\frac{7.6692^8}{8!\pi r_{eff}^2} e^{-7.6692(r/r_{eff})^{1/4}}
\end{equation} 

with $r$ the radial distance as measured from the centre of the galaxy, and $L_{tot}$ the extrapolated total galaxy light. Spherical symmetry was assumed for the deVaucoleurs galaxies. The $r^{1/4}$-profile is very steep at small radii, and thus these types of galaxies are detected down to a fainter limiting magnitude with respect to disk galaxies. The total light is however distributed over a more extended region-- and therefore, the amount of flux lost in the low-surface brightness wings is larger -- than for an exponential profile. 

The idealized galaxies were convolved with the ACS PSF before being added to the NICP12 and HUDF images. The PSF was derived from a relatively bright but unsaturated star in the field. Tests with a PSF produced with TinyTim \citep{kri04}, downgraded to the measured FWHM of the observed PSF, were also performed, and produced similar results. Poisson noise from the galaxy counts was not added to the simulated galaxies, since at the faint magnitudes of this study the sky background is the dominant source of Poisson noise. Tests were nonetheless performed, which confirmed that the galaxy shot noise had no impact on the results.

\subsection{Dimmed-Galaxies Simulations}
\label{sec:dimdetails}

The idealized-galaxies simulations allow us to relate the intrinsic properties of galaxies to how these would be observed with our instrumental setup. They suffer however from the limitation of assuming perfectly smooth light distributions, as described by the adopted analytical descriptions. This is clearly not the case for real galaxies at any epoch, and certainly not for the $UV$ light emerging from the star forming $z\sim5$ galaxy population.

In order to have a benchmark to gauge the incompleteness in our samples on the basis of more realistic assumptions concerning the galaxy light distributions, we performed an additional set of simulations, entirely based on the galaxies that were detected in the NICP12 and HUDF fields. Specifically, we dimmed the ACS images by specific amounts, and added these dimmed images to the corresponding original frames. Offsets of a few pixels were introduced before co-adding the dimmed and original images. These offsets were chosen large enough to ensure that a given original galaxy was not 'self-blended' to its dimmed version in the subsequent SExtractor run. Since this procedure adds additional noise to the images, the co-added images were appropriately rescaled in order that galaxies of any given magnitude will have the same S/N values in the original and the final image. More specifically: 

\begin{enumerate}
\item A copy of the original image was divided by a factor $f$;
\item This $f$-scaled version of the data was shifted by a few pixels relative to the original parent frame and added to the latter;
\item The resulting composite image was divided by $\sqrt{1+\left(1/f\right)^2}$, to retain the S/N vs. magnitude relation of the original image;
\item The rms map of this final "dimmed image" was obtained from the rms map of the original image ($\sigma_{orig}$) and the shifted image ($\sigma_{shift}$) by computing:
\[
\mbox{rms}=\frac{\sqrt{\sigma_{shift}^2/f^2+\sigma_{orig}^2}}{\sqrt{1+1/f^2}}
\]
\end{enumerate}

The procedure leads to an effective dimming factor for the galaxies' fluxes of $n_{dim}=f\cdot\sqrt{1+\left(1/f\right)^2} = \sqrt{1+f^2}$. 
We performed four separate experiments with four different shifts and dimming factors $n_{dim} = \sqrt{2},2,4,6$. For each experiment we shifted and dimmed the \zF, \iF, and \vFilter images by the same amount and reran SExtractor with the same parameters as for the compilation of the original catalogues. 
We used these "dimmed-galaxies" simulations to quantify, in an independent and complementary way relative to the "idealized-galaxies" simulations discussed above, the completeness of our catalogs, as well as the scatter in measured magnitudes and colors, as a function of galaxy luminosity.

Note that these simulations differ conceptually from the 'cloning technique' adopted by \citet{bouw05} as we add dimmed versions of the whole ACS images to the original tiles rather than inserting modified cut-outs of a subset of galaxies.


\end{document}